\pdfoutput=1
\documentclass[a4paper,11pt]{JINST}

\usepackage{textcomp}
\usepackage{amsmath}
\usepackage{lineno}
\newcommand{\cfour}{$^{14}$C}
\newcommand{\radon}{$^{222}$Rn}

\title{Borexino calibrations: Hardware, Methods, and Results}

\author{
H.~Back$^{3}$, 
G.~Bellini$^{1}$, 
J.~Benziger$^{6}$, 
D.~Bick$^{17}$,
G.~Bonfini$^{2}$, 
D.~Bravo$^{3}$, 
M.~Buizza~Avanzini$^{5}$,
 B.~Caccianiga$^{1}$, 
L.~Cadonati$^{16}$, 
F.~Calaprice$^{4}$, 
C.~Carraro$^{7}$, 
P.~Cavalcante$^{2}$, 
A.~Chavarria$^{4}$, 
A.~Chepurnov$^{18}$,
D.~D{\textquoteright}Angelo$^{1}$, 
S.~Davini$^{7,20}$, 
A.~Derbin$^{9}$, 
A.~Etenko$^{10}$, 
F. von Feilitzsch$^{8}$,
G.~Fernandes$^{7}$, 
K.~Fomenko$^{11}$, 
D.~Franco$^{5}$, 
C.~Galbiati$^{4}$, 
S.~Gazzana$^{2}$, 
C.~Ghiano$^{2}$,
M.~Giammarchi$^{1}$, 
M.~Goeger-Neff$^{8}$, 
A.~Goretti$^{4}$, 
L.~Grandi$^{4}$, 
E.~Guardincerri$^{7}$, 
S.~Hardy$^{3}$, 
Aldo Ianni$^{2}$, 
Andrea Ianni$^{4}$, 
A.~Kayunov$^{9}$, 
S.~Kidner$^{3}$,
V.~Kobychev$^{19}$, 
D.~Korablev$^{11}$, 
G.~Korga$^{13,20}$, 
Y.~Koshio$^{2}$, 
D.~Kryn$^{5}$, 
M.~Laubenstein$^{2}$, 
T.~Lewke$^{8}$, 
E.~Litvinovich$^{10}$, 
B.~Loer$^{4}$, 
F.~Lombardi$^{2}$, 
P.~Lombardi$^{1}$, 
L.~Ludhova$^{1}$, 
I.~Machulin$^{10}$, 
S.~Manecki$^{3}$, 
W.~Maneschg$^{12}$, 
G.~Manuzio$^{7}$, 
Q.~Meindl$^{8}$, 
E.~Meroni$^{1}$, 
L.~Miramonti$^{1}$, 
M.~Misiaszek$^{14}$, 
D.~Montanari$^{2,4}$, 
P.~Mosteiro$^{4}$,
V.~Muratova$^{9}$, 
L.~Oberauer$^{8}$, 
M.~Obolensky$^{5}$, 
F.~Ortica$^{15}$,
K.~Otis$^{16}$,  
M.~Pallavicini$^{7}$, 
L.~Papp$^{13,3}$, 
L.~Perasso$^{7}$, 
S.~Perasso$^{7}$, 
A.~Pocar$^{16}$, 
R.S.~Raghavan$^{3}$, 
G.~Ranucci$^{1}$, 
A.~Razeto$^{2}$, 
A.~Re$^{1}$, 
A.~Romani$^{15}$, 
N.~Rossi$^{2}$,
D.~Rountree$^{3}$,
A.~Sabelnikov$^{10}$, 
R.~Saldanha$^{4}$, 
C.~Salvo$^{7}$, 
S.~Sch\"onert$^{8}$, 
H.~Simgen$^{12}$, 
M.~Skorokhvatov$^{10}$, 
O.~Smirnov$^{11}$, 
A.~Sotnikov$^{11}$, 
S.~Sukhotin$^{10}$, 
Y.~Suvorov$^{10,21}$, 
R.~Tartaglia$^{2}$, 
G.~Testera$^{7}$, 
D.~Vignaud$^{5}$, 
R.B.~Vogelaar$^{3}$, 
J.~Winter$^{8}$, 
M.~Wojcik$^{14}$, 
A.~Wright$^{4}$, 
M.~Wurm$^{17}$, 
J.~Xu$^{4}$, 
O.~Zaimidoroga$^{11}$, 
S.~Zavatarelli$^{7}$,
G.~Zuzel$^{14}$\\
\center{(Borexino Collaboration) \footnote{E-mail: {\it
spokeperson@borex.lngs.infn.it}}}\\

$^1$Dipartimento di Fisica, Universit\`a di Milano and INFN Milano, via Celoria 16, \\ I-20133 Milano, Italy

$^2$Laboratori Nazionali del Gran Sasso, SS 17bis Km 18+910, I-67010 Assergi (AQ), Italy

$^3$Physics Department, Robeson Hall, Virginia Polytechnic Institute and State University, Blacksburg, VA 24061-0435, USA 

$^4$Department of Physics, Princeton University, Jadwin Hall, Washington Road, Princeton, \\ NJ 08544-0708, USA

$^5$Astroparticule et Cosmologie APC, 10 rue Alice Domon et L\'eonie Duquet, \\ 75205 Paris cedex 13, France

$^6$Department of Chemical Engineering, Princeton University, Engineering Quadrangle, Princeton, NJ 08544-5263, USA

$^7$Dipartimento di Fisica, Universit\`a di Genova and INFN Genova, via Dodecaneso 33, \\ I-16146 Genova, Italy

$^8$Technische Universit\"at M\"unchen, James Franck Strasse E15, D-85747 Garching, Germany

$^9$St. Petersburg Nuclear Physics Institute, Gatchina, Russianon

$^{10}$NRC Kurchatov Institute, Kurchatov Sq. 1, 123182 Moscow, Russia

$^{11}$JINR, Joliot Curie str. 6, 141980 Dubna, Russia

$^{12}$Max-Planck-Institut f\"ur Kernphysik, Saupfercheckweg 1, D-69117 Heidelberg, Germany

$^{13}$KFKI-RMKI, 1121 Budapest, Hungary

$^{14}$Institute of Physics, Jagiellonian University, ul. Reymonta 4, PL-30059 Krakow, Poland

$^{15}$Dipartimento di Chimica, Universit\`a di Perugia and INFN Perugia, via Elce di Sotto 8, I-06123 Perugia, Italy

$^{16}$Physics Department, University of Massachusetts, Amherst, MA 01003, USA

$^{17}$Institut f\"ur Experimentalphysik, Universit\"at Hamburg, 22761 Hamburg, Germany

$^{18}$Physics Department, Lomonosov Moscow State University, 119899, Moscow, Russia

$^{19}$Institute for Nuclear Research, 03680 Kiev, Ukraine,

$^{20}$Physics Department, University of Houston, Houston, TX 77204, USA. 

$^{21} $Physics and Astronomy Department,  University of California Los
Angeles, Los Angeles, CA 90095, USA.

}

\abstract{Borexino was the first experiment to detect solar neutrinos 
in real-time in the sub-MeV region. In order to achieve
high precision in the determination of neutrino rates, the detector design
includes an internal and an  external calibration system. 
This paper describes both calibration systems and the calibration campaigns 
that were carried out in the period between 2008 and 2011.
We  discuss some of the results and show that the calibration procedures 
preserved the radiopurity of the scintillator. 
The calibrations provided a detailed understanding of the detector response 
and led to a significant reduction of the systematic uncertainties in the 
Borexino measurements.}


\begin{document}


\section{Introduction}
\label{intro}

Borexino is a large volume liquid scintillator detector located at the 
Laboratori Nazionali del Gran Sasso (LNGS), Italy.
The main goal of this experiment is the study of the low energy part of 
the solar neutrino spectrum in the sub-MeV region, in particular the monochromatic $^7$Be neutrinos. 
Borexino succeeded to perform the first real-time detection of $^7$Be neutrinos \cite{borex_be7_1}
and measured the $^7$Be neutrino flux and its day/night asymmetry \cite{borex_be7_2, borex_be7_3, borex_be7_daynight} with a high precision. 
Furthermore, a detailed solar neutrino spectroscopic measurement of the solar $pep$ and $^8$B neutrinos was carried out and the most 
most stringent limit on the CNO neutrino rate was determined \cite{borex_b8, borex_pep}. 
Borexino also proved to be a clean anti-neutrino detector that was able to observe geo-neutrinos \cite{borex_geonu} and to set new limits on anti-neutrino fluxes from the Sun and 
other unknown sources \cite{borex_antinu}.
Finally, the extremely  clean environment of the Borexino detector allowed to search for rare or even forbidden processes like Pauli violating
transitions \cite{borex_Pauli}, or for axions produced in the Sun \cite{borex_axion}.
  
The key requirements for the success of Borexino are the 
radiopurity of its scintillator and the complete knowledge of the
detector's response. The latter is needed for several reasons: 
since the neutrino interaction rate is determined from a fit to the 
scattered electron energy spectrum, it is necessary to determine 
the energy scale to high precision over a broad energy range 
(between 0.1 and 10\,MeV depending on the analysis).
The large dimensions of the detector require a 
careful mapping of its energy response in 
different positions within the scintillating volume. Another important 
issue is the testing and tuning of $\alpha$/$\beta$
discrimination techniques. Finally, most analyses rely on the offline software-cut selection 
of scintillator subvolumes (so-called fiducial volumes (FV)) that optimize the
signal-to-background ratio. Thus, it is critical as well to validate the position 
reconstruction algorithm, as this reduces systematic uncertainties in the 
evaluation of the target mass.
All tasks can be accomplished through calibrations using 
sources of different types ($\alpha$, $\beta$, $\gamma$ and neutron
emitters; laser source) that cover different energy regions.

During the first period of data-acquisition from May 2007 until 
October 2008, no radioactive sources were
inserted into the detector due to the risk of contamination of the highly radiopure scintillator. 
At that time, contaminants already present in the detector
such as $^{14}$C and $^{222 }$Rn or the cosmogenically produced $^{11}$C 
were used as calibration candles allowing the first measurement of the $^7$Be neutrino rate
with a total uncertainty budget 
of $\simeq$~10\% \cite{borex_be7_1, borex_be7_2}. To improve 
this result, several calibration campaigns were performed starting from 
October 2008. Sources of different
types were deployed in different positions within the detector. 
The calibrations provided a detailed understanding of the detector 
response, which, in case of the $^7$Be neutrino rate made possible to reduce 
the systematic uncertainty from 8.5\%  to less than 2\% \cite{borex_be7_3, borex_be7_daynight}.

This paper is devoted to the description of the Borexino 
calibration systems and  calibration campaigns.
Two separate sub-systems are presented. The first one is the 
internal calibration system which was designed for the insertion
of radioactive sources at different positions within the scintillator volume. 
The second one
is the external calibration system that was used to deploy $\gamma$ sources 
into the buffer region
around the stainless steel sphere. Both systems were carefully designed to comply with the 
stringent radiopurity specifications of the Borexino experiment.

Section~\ref{BX} briefly describes the Borexino detector.
Section~\ref{sec:int} is devoted to 
the internal calibration system: hardware,  sources and
calibration procedures are described.
Similarly, Section~\ref{sec:ext_calibr_system} is devoted to the description of the 
external calibration system.
Several relevant examples of results obtained from the
calibrations are presented in Section \ref{CR}. 
These results show that the calibration 
systems worked properly and that the 
calibrations significantly improved Borexino's physics results.

\section{The Borexino Detector}
\label{BX}
\begin{figure}[htb]
\centering
\includegraphics[width=0.8\columnwidth]{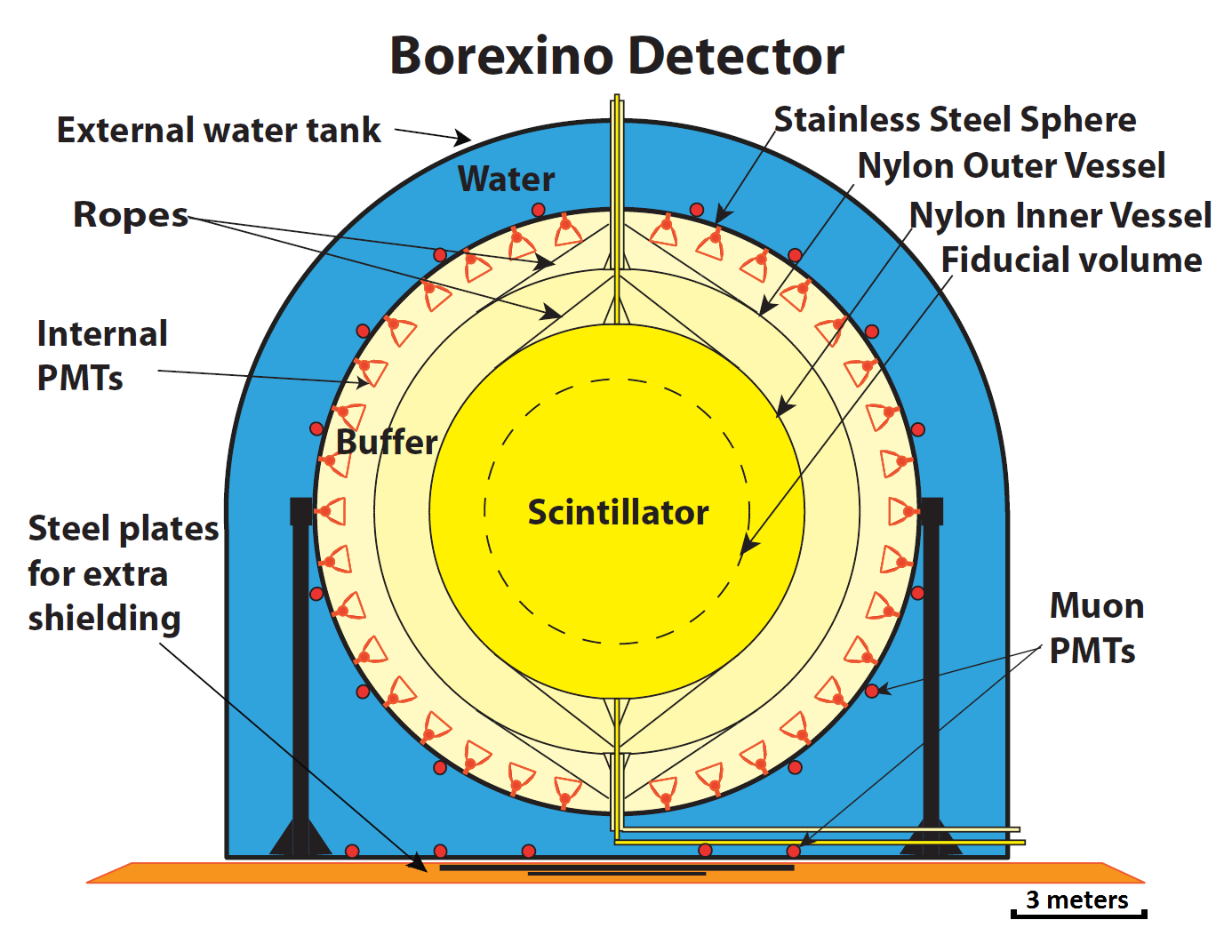}
\caption{Schematic view of the Borexino detector.}
\label{fig:BX}
\end{figure}

Borexino is located deep underground at 3,800 meters of water equivalent, 
in the Hall C of the Laboratori Nazionali
del Gran Sasso (LNGS). At this depth the cosmic muon flux is reduced by a factor $\sim$10$^6$.
Borexino detects neutrinos through the elastic scattering interaction with electrons 
in the organic liquid scintillator. Anti-neutrinos are detected via the
inverse neutron beta-decay.
The detector design follows the concept 
of graded shielding, in which concentric material layers of increasing radiopurity
shield the innermost ultra-pure core of the experiment.
A schematic view of the Borexino structure is depicted in Figure~\ref{fig:BX}. 
The detection medium is an organic liquid scintillator 
consisting of 278\,tons (315\,m$^3$) of pseudocumene (1,2,4-trimethylbenzene) 
as a solvent
doped with 1.5\,g/l PPO (2,5-diphenyloxazole) as a solute. 
The scintillator needs to be exceptionally radiopure
to be capable of detecting the feeble
neutrino signal (which amounts to a few tens of counts/(day$\times$100\,ton). 
A dedicated purification strategy
including distillation and sparging with ultrapure nitrogen has been 
developed during 15 years of R\&D studies and was successful
in reducing background down to the required levels. As an example, the 
residual $^{238}$U and $^{232}$Th concentrations in the scintillator are 
(1.6$\pm$0.1)$\times$10$^{-17}$g/g and 
(6.8$\pm$1.5)$\times$10$^{-18}$g/g, respectively: these unprecedented 
levels of radiopurity even exceed the design goals.
The scintillator is contained in a spherical nylon 
balloon of radius $R$=4.25\,m. This inner
vessel (IV) is viewed by 2214 photomultiplier tubes (PMTs) mounted on a 
concentric stainless steel sphere (SSS) of radius
$R$=6.85\,m.
Between the SSS and the IV an ultra-pure buffer liquid (pseudocumene + DMP
 to quench scintillation light) effectively 
minimizes the
diffusion of radioactive particles and the penetration of external 
radiation from the PMTs, 
1843 light concentrators (LCs), and the SSS. A spherical nylon barrier (Outer Vessel, OV) 
concentric with the IV prevents Radon emanated from the 
materials on the SSS to reach the scintillating core of the detector.
Further shielding against radioactivity from the surrounding rocks 
is provided by 2100\,tons of ultra-pure water in a domed steel tank 16.9\,m in height and
18\,m in diameter. This outer detector (OD) embeds the SSS containing the inner 
detector (ID). The OD is also equipped with PMTs to detect 
Cherenkov light emitted by residual cosmic muons crossing the water. 
More details about the Borexino detector and about its purification 
facilities are presented in \cite{borex_liquid, borex_NIM}.

In this context, calibrations are an important and delicate issue: in fact, 
the necessity to preserve the
detector radiopurity poses severe constraints on the system 
design and raises several problems from the operational point of view.
We show in this paper that these difficulties have been overcome
and that the system has worked properly, becoming an important element for the success of the Borexino experiment.

\section{Internal Source Calibration}
\label{sec:int}

\subsection{Hardware}

The  internal calibration system of the Borexino detector has a complex
structure that consists of two sub-systems. The first one is the 
calibration source deployment system 
(Section~\ref{sec:int_depl}) used to deploy  radioactive or laser sources
into the desired location within the scintillator. 
The hardware design incorporates solutions to several obstacles 
related to the detector requirements and provides operational comfort 
during calibration campaigns. For instance, even though all the operations can be performed 
by only one operator, the hardware design allows for two people to share the same task.
The second sub-system is the  source location system
(Section~\ref{sec:cal_loc_sys}) used to determine the source reference 
position with a precision of better than 1\,cm.

\subsubsection{Source Insertion System} 
\label{sec:int_depl}

\begin{figure}[htb] 
\centering
\includegraphics[width=0.55\textwidth]{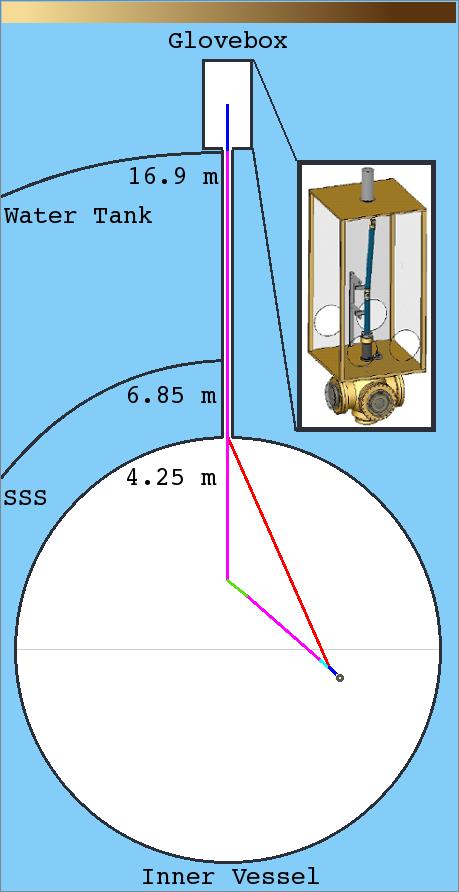}
\caption{Schematic view of the source deployment system: the main figure
shows the glovebox and Inner Vessel, together with the pipe connecting them.
The inset to the right shows a zoomed view of the glovebox. 
 In order to reach the desired location, the hinge is positioned and the 
 system is lowered vertically into the detector.  In the next step, 
 the tether tube (red line) is 
  withdrawn until the hinge bends to the chosen angle. To establish 
  the azimuthal position in $\phi$ the rods are simply rotated $\pm 180^\circ$ -- 
  procedural limitations prohibit rotation by more than $180^\circ$ in either 
  direction.  To retract the source, the procedure is reversed.}
\label{fig:source_deploy_sequence}
\end{figure}

Due to the detector's specific design the only access to the scintillator volume 
is through a 4''
diameter pipe connecting the IV to a gate valve on top of the 
water tank --  a vertical distance 
of six meters (see Figure ~\ref{fig:source_deploy_sequence}).  
The source deployment system itself is made up of a series of interconnecting
hollow rods, assembled into an arm that can be bent up to $90^\circ$ once 
inside the detector.
All operations are performed through a glovebox  installed in a Class 100 
clean-room atop the detector; 
an automated process-control system monitors gas pressures and 
flow rates inside the glovebox and the connecting equipment.  

\paragraph{Hardware}
\label{sec:int_sys}

\begin{figure}[htb]
\centering
\includegraphics[width=0.6\textwidth]{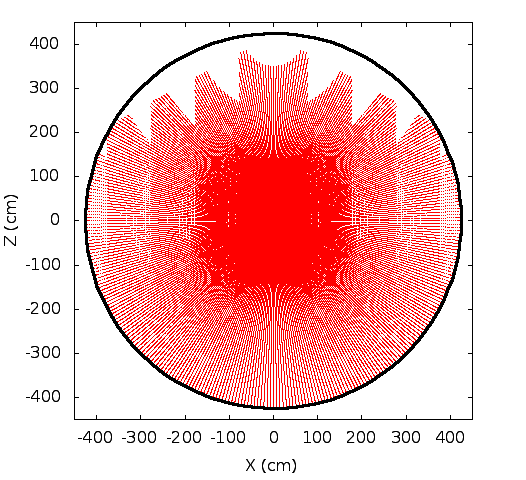}
\caption{Schematic view of areas reachable with the Borexino source deployment system.  
The red points represent the mesh of positions where a source can be deployed
within the scintillator volume. The black line represents the nominal position of the inner vessel.}
\label{fig:go_nogo_areas}
 \end{figure}

Each of the insertion rods (3.8\,cm$\times$100\,cm) is equipped with special 
couplers at both ends.
The rods also contain a ballast wire which is sized to make the rods almost
neutrally buoyant when immersed in the Borexino scintillator.
A hinge in the lever arm prevents any motion over $90^\circ$, and 
 can be used in place of any normal rod 
in order to facilitate longer lever arms. Figure \ref{fig:go_nogo_areas} 
shows a map of areas that can be reached using this system.

\begin{figure}[htb] 
\centering
\includegraphics[width=0.6\textwidth]{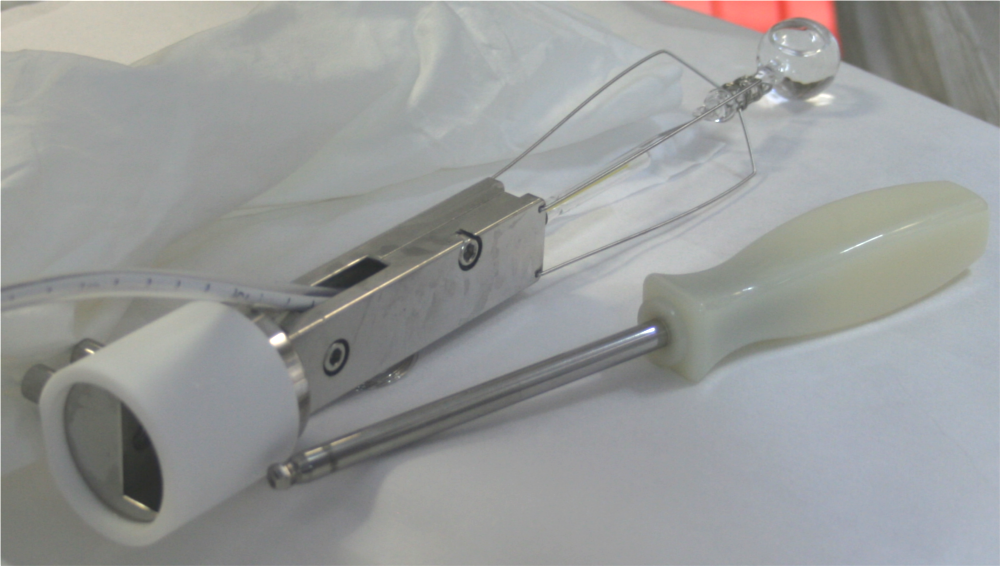}
\caption{The special coupler used to attach the insertion rods to a calibration source and tether 
(entering the coupler midway down its body). A thick collar on the left hand side prevents the coupler 
from being drawn into the sliding seal. The four spring-steel support wires to which the source is 
attached are visible at the end of the coupler.}
\label{fig:source_coupler}
\end{figure}

A 1/4'' diameter flexible Teflon tether tube enters 
the detector alongside the rods.
The tether is fixed at the end of the last rod, the one holding the source, and
is used to adjust the hinge angle to the desired position (for details see caption of
Figure \ref{fig:source_deploy_sequence}). The tether also carries a
fiber-optic cable that is used in determining source position (Section \ref{sec:cal_loc_sys}). 
Both rods and tether are attached to a special source-coupler 
which holds the calibration source
at its end with four spring-steel wires; Figure \ref{fig:source_coupler} 
shows the assembled configuration. 

The glovebox, which also stores all of the components, remains under 
continued flow of low argon/krypton nitrogen (LAKN$_2$: 0.01\,ppm Ar, 0.02\,ppt Kr) at a pressure of 2-3\,mbar. 
An oxygen monitor is installed in the clean-room for safety reasons,
since, in case of a leak, nitrogen could exit the glove-box and
 displace oxygen in the clean-room. 
A leak in the glove-box would be dangerous for the detector as well, because of
Rn, Kr, Ar, and other contaminants feeding back into the scintillator.

\begin{figure}[htb] 
\centering
\includegraphics[width=0.4\textwidth]{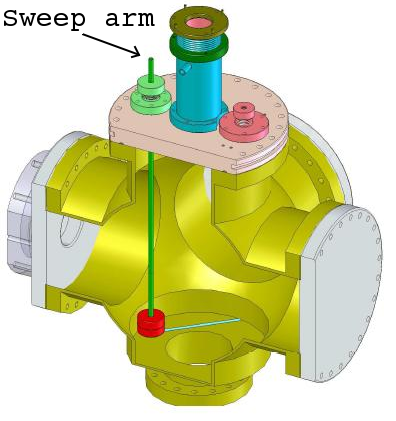}
\caption{A six-way vacuum cross that was mounted between the glovebox and the gate valve. An additional 
sweep arm is used to verify the position of the rods during the source extraction.}
\label{fig:sweep-arm}
\end{figure} 

The fluid handling system in the Borexino detector \cite{borex_liquid} requires that a slight
(100\,mbar) overpressure be maintained in the IV. In order to maintain this
pressure while deploying sources, a 6-way cross shown in Figure \ref{fig:sweep-arm} with a sliding seal that
mated around the source deployment rods at the top, was located between
the glovebox and the gate valve (a diagram showing the location of the cross with respect 
to the glovebox is presented in the inset of Figure \ref{fig:source_deploy_sequence}). 
After a source was mounted and the first
deployment rod enclosed within the sliding seal, the cross was brought to
the necessary pressure using a flow of LAKN before the gate valve to the
IV was opened.

\paragraph{System Cleanliness}
\label{sec:int_clean}

As cleanliness is the driving constraint for the system described so far, all components entering the detector 
(insertion rods, couplers, tether tube, and sources) have to meet the same cleanliness requirements adopted for the filling stations.  
The electropolished rods and couplers were placed, four at a time, into a vessel
 connected to the Borexino cleaning module
for detergent cleaning as described in \cite{borex_liquid}.  
When particulate counting of the rinse water indicated a cleanliness level of Class 30 or better as defined by
MIL-STD 1246C \cite{MilStd1246C}, the components were unloaded into a clean-room for assembly.
The cross was also cleaned in the same manner because the scintillator could potentially reach this area if control of the liquid 
level was lost. Due to the small size of the tether and sources, these components were cleaned in ultrasonic baths prior to their usage. 
As shown in Section \ref{sec:SC} the effort devoted to cleaning the system allowed a successful calibration of
the Borexino detector while preserving its excellent scintillator radiopurity.

\paragraph{Control Software}
\label{sec:int_control-software}

Control over the system is largely performed by a custom-written program that provides a graphical-user-interface, 
manual and automated control over solenoid-operated components, visual and audible 
alarms, and data logging capability. 


\subsubsection{Source Location System}  
\label{sec:cal_loc_sys}

A task of paramount importance in Borexino analyses is the definition of
the fiducial volume (FV), which is directly related to the determination of the 
event position.  Thus, one objective of the calibration campaigns was to study
the performance of the software algorithm used to estimate the positions of
data events by comparing the source locations determined using the reconstruction
software with the true source locations. The method is ultimately limited by the uncertainty on the true position 
of the source. Achieving an uncertainty on the fiducial volume  determination 
at the 1\% level requires a precision of $\sim$1\,cm on the source position determination. 

The source location system consists of seven consumer grade digital cameras, six of which are on orthogonal axes.
Camera 7 was installed close to the top of the detector in order to monitor for trapped gas bubbles during filling.
The true location of a source can be determined in the following way: 
a laser-illuminated diffuser ball, attached close to the source, 
is flashed while the  CCD cameras take pictures simultaneously. It is worth noticing that the color of the laser 
was selected intentionally, red light is visible in the spectrum of the cameras, but it is not harmful to the PMTs. 
In such a configuration, it was safe to leave the high-voltage turned on while the pictures were taken.

\begin{figure}[htb]
\centering
\includegraphics[width=0.7\textwidth]{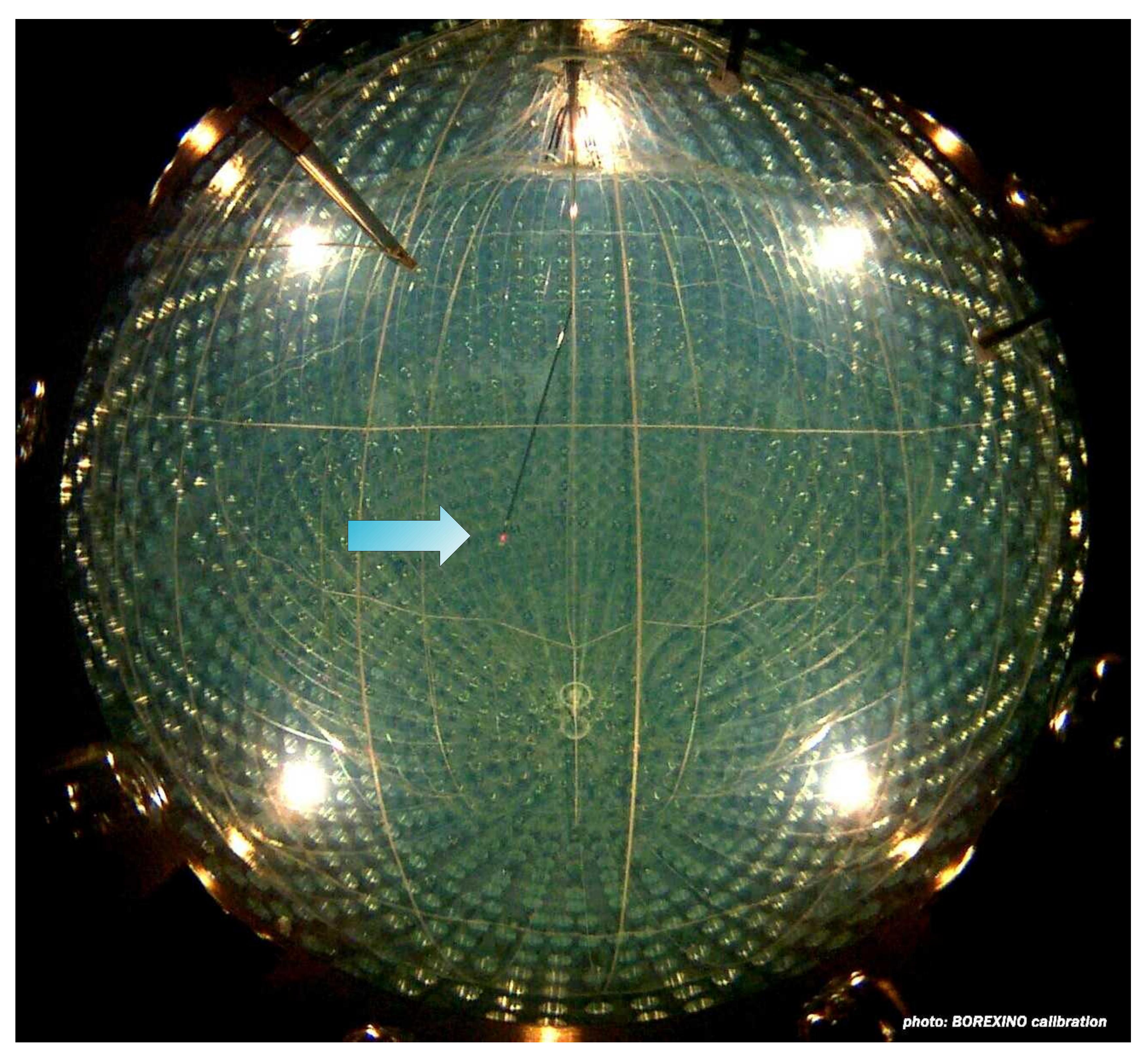}
\caption{CCD image of the Borexino inner detector showing the insertion arm during an 
internal calibration. The diffuser for the location of the
source is visible in red close to the center of the detector. To take this
picture lights were turned on for illustration purposes. During normal 
calibration activities they would be off.}
\label{fig:photo_diffuser_source}
\end{figure}

An example is given in Figure \ref{fig:photo_diffuser_source}. The position of the diffuser and of the source is determined
via triangulation of all pictures. Although finding a point in three dimensions requires only two cameras,
all seven are used to check for self-consistency and to increase the resolution of the system. 

The reference coordinate system adopted throughout this paper
has the origin in the SSS center and the z-axis along the vertical.

\paragraph{CCD Cameras and Computer Interface}

\begin{figure}[htb]
\centering
\includegraphics[width=0.7\textwidth]{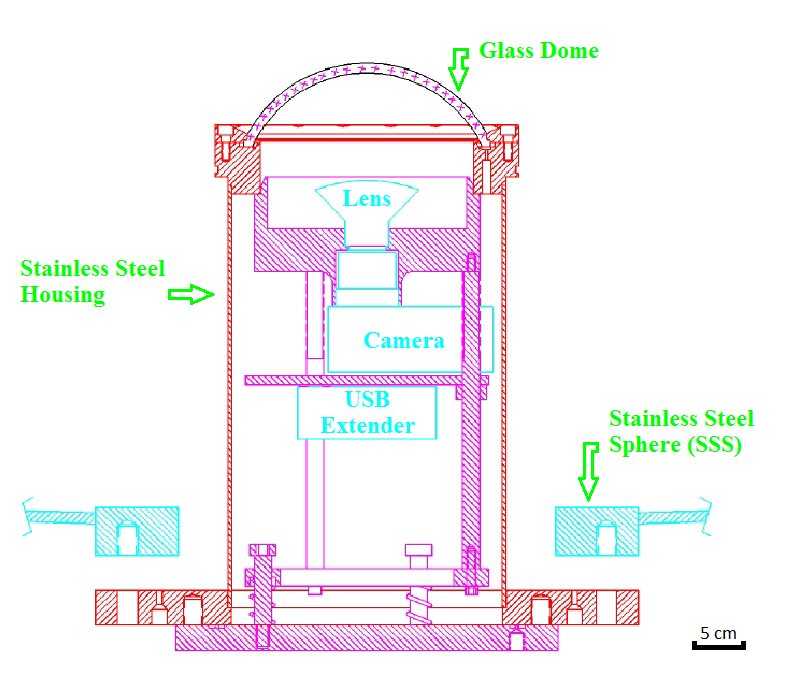}
\caption{Assembly drawing of a camera canister.  
A fisheye lens was added to each assembly to yield a $183^\circ$ field of view for each camera. 
A USB extender in each housing transmits the camera's USB signal over CAT5 cable; an analogous unit
at the receiving end transforms the signal back to USB for computer read-out.}
\label{fig:cam_geom}
\end{figure}

The CCD camera system employs Kodak DC290 $2.4$ megapixel consumer grade digital cameras, 
each equipped with a Nikon FC-E8 fisheye lens.  The fisheye lens allows the entire IV to be viewed by 
expanding each camera's field of view to $183^\circ$.  The camera + lens system is mounted in a  
housing on the SSS, with a glass underwater photography dome on the front side. Figure \ref{fig:cam_geom} 
illustrates the camera housing and the mounting design.  

The cameras are controlled over an interface with the camera control box.
This box houses relays and circuitry to ensure synchronized photo-taking. Additionally, the camera housing can be flushed with
nitrogen in order to remove unwanted water residuals or scintillator vapors.

There are also four pin-hole LEDs present within each of the camera housings to provide fixed reference points 
for evaluating time stability of the location system. In addition, the pin-hole lights are visible to the same
camera they are mounted with, and are used to correct motions in that camera's lens. Including the light diffuser
mounted above the source, there are a total of 29 red light sources in fixed locations
about the SSS that are flashed when a photo is taken to provide reference positions. Tests were performed with high voltage of all 
PMTs turned on, and an increase in the dark rate for illuminations as long as two seconds was not found.  
During normal calibration operations, photos are only taken when the DAQ has been stopped (but with the PMT high
voltage still on).

A total of eight 50\,W quartz-halogen lights, split into two independently wired banks, are installed in 
each camera housing to illuminate the detector for vessel monitoring.  
As a safety precaution, the lights are enabled with a key switch; the switch also disables the PMT high voltage. 
This interlock method ensures that the PMTs will never be exposed to the intense halogen light while the high voltage is on. 
The camera control box also contains a timer circuit which shuts the lights off after two seconds to prevent 
excessive heating of the housing.

The source location system must comply with the stringent radiopurity 
requirements of the Borexino apparatus.
In particular, since the system components are installed on the SSS, 
we have ensured that their activity is low 
in comparison with the other sources of external background located
 at the same
radial distance, like 
the PMTs, the light concentrators and the SSS itself.  
The $^{238}$U and $^{232}$Th concentration in the location system 
was measured to be negligible in comparison with that
of the PMTs \cite{HB_Dissertation}. 
The contamination of $^{40}$K was found to be higher than in the 
PMTs by about 85\%. This was considered acceptable, since it 
accounts for a relatively small fraction (9\%) of the total 
external background rate 
in Borexino.

\paragraph{Image Reconstruction}
\label{sec:image_recon}

Triangulation of the diffuser is achieved by projecting a ray in space from each camera to the diffuser,
and then finding the intersection of the seven rays. In order to find a ray from one camera, it would be necessary 
to track the light through the camera's lens system, which not only requires detailed information about 
the lenses, but also a precise measurement of the camera's mounting position and direction. 
To overcome these difficulties, the image reconstruction process employs a transformation method that 
corrects for lens distortions \cite{lens_corrections}, camera mounting uncertainties, and performs the correct 3D mapping of the 
detector onto the CCD. The transformation process uses thirteen parameters which are determined by fitting the
well-known spatial coordinates of the PMTs in images of the detector taken with the lights on, as will be
described later.  

\paragraph{Camera Calibration}
\label{sec:calibration-of-ccd}

During normal operation, the camera system uses the parameters mentioned in 
the previous paragraph coupled with an image to
determine the position of an object in that picture.  However, this process requires 
\emph{a priori} knowledge of the calibration
parameters.  Obtaining the calibration parameters can be accomplished by effectively 
running the method in reverse, and using
the \emph{known} positions of the inner detector PMTs to determine the \emph{unknown} 
calibration parameters.  

The calibration process begins by creating an idealized picture -- ellipses to represent 
the edges of the PMT light concentrators and
$\mu$-metal shielding -- of what each camera would see and then overlaying this idealized 
picture onto an actual photo taken with the
cameras.  A user can then drag the overlay image until a given PMT on the overlay lines up
 with one in the actual image; the point is
then saved, and the process is repeated for 100 or more points.  At the end of this process, 
a $\chi^2$ minimization is performed to
determine the calibration parameters that would produce a corrected image lining up with the overlay.  
At the conclusion of this
process, the thirteen parameters resulting from the calibration allow the corrected images to be 
treated as if they were taken by an
ideal camera, vastly simplifying ray-tracing.

\paragraph{Position uncertainty}
\label{sec:uncertainty}

In order to estimate the uncertainty on the determination of the source position by the source location system, 
we performed several tests. First, we verified that the detector center as determined by the cameras coincides 
with the nominal center, derived assuming a perfectly spherical SSS.
This was done by verifying that each camera ray that traverses the center of the SSS at (0,0,0)\,cm also points at the center of the opposite camera
across the detector. Next, using a fixed reference point in the glovebox, we obtained the absolute scale
by comparing the reconstructed LED position with the total length of the rods for different locations along the z-axis
(maximum uncertainty on the length of the rods is $\pm$0.2\,cm). In this way we derived an uncertainty for the determination 
of the z position of $\sim$\,0.6\,cm. Even though this test was performed only on the z axis, it can
be safely extrapolated to other positions, since the cameras are not located in any special orientation with respect to 
the z-axis.

\begin{figure}[htb]
\centering
\includegraphics[width=0.5\textwidth]{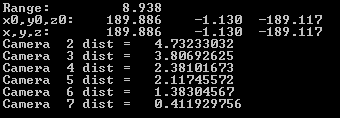}
\caption{A view from the software window indicating the reconstructed source position (in cm) and 
shortest distances to each of the rays (in pixels). An approximate conversion at the center of the detector is about 1\,cm = 1\,pixel.}
\label{fig:rayd} 
\end{figure}

As an additional check, the camera reconstruction software gives us the 
opportunity to verify how far each of the rays was from the 
calculated position. 
As an example, Figure~\ref{fig:rayd} shows the reconstructed position for one 
of the sources with calculated distances to each of the camera-rays. 
Six out of seven installed cameras identified this source correctly however, 
cameras 2 and 3 could be eliminated due to higher than expected distance, 
4.7 and 3.8 pixels respectively\footnote{The view from camera 1
was possibly shadowed by the calibration rods and hence, was not included in the reconstruction at all.}.
The position was afterwards re-calculated
without including these two cameras. This procedure has been adopted for 
the determination of all source positions and improved the reliability of
the result.
It was also used to investigate possible systematics of the location system  
connected to the source position:
no directionality or radial dependence was identified.

\subsection{Calibration Sources}
\label{sec:sources}

As shown in Table \ref{tab:sources}, sources of different types were needed to cover the energy region 
of interest for Borexino and to investigate the scintillator response to different ionizing particle types. A detailed description
of each of the sources is presented in the following Sections \ref{sec:rn222-c14-sources} to \ref{sec:laser-source}.

\begin{table*}
\begin{center}
\begin{tabular}{cccccc}
\hline
\hline
Source & Type &E [MeV] &Position & Motivations &Campaign\\
\hline
$^{57}$Co & $\gamma$ &0.122 &in IV volume  & Energy scale  & IV\\
$^{139}$Ce & $\gamma$&0.165 &in IV volume  & Energy scale & IV\\
$^{203}$Hg & $\gamma$&0.279 &in IV volume  & Energy scale & III\\
$^{85}$Sr & $\gamma$ &0.514 &z-axis + sphere R=3\,m  & Energy scale + FV& III,IV \\
$^{54}$Mn & $\gamma$ &0.834 &along z-axis  & Energy scale & III\\
$^{65}$Zn & $\gamma$ &1.115 &along z-axis & Energy scale& III \\
$^{60}$Co & $\gamma$ &1.173, 1.332&along z-axis & Energy scale & III\\
$^{40}$K & $\gamma$ &1.460 &along z-axis& Energy scale & III\\
\hline
$^{222}$Rn+$^{14}$C  & $\beta$,$\gamma$&0-3.20&in IV volume& FV+uniformity & I-IV\\
           & $\alpha$&5.5, 6.0, 7.4&in IV volume& FV+uniformity&  \\
$^{241}$Am$^{9}$Be & n & 0-9&sphere R=4\,m& Energy scale + FV& II-IV \\ 
394 nm laser & light &-&center&PMT equalization& IV\\
\hline
\hline
\end{tabular}
\caption{Radioactive sources used during the Borexino  internal calibration 
campaigns. The radionuclides, energies and emitted particle types are 
shown in the first three columns. The fourth column indicates the 
positions where the sources were deployed within the scintillator. 
The main purposes for the individual source measurements are summarized in the
fifth column.
The last column indicates in
which campaign the sources have been deployed: I (October 2008), II (January
2009), III (June 2009) and IV (Jul 2009), see text for more details.}
\label{tab:sources}
\end{center}
\end{table*}

\begin{figure}[htb]
\centering
\includegraphics[width=0.5\textwidth]{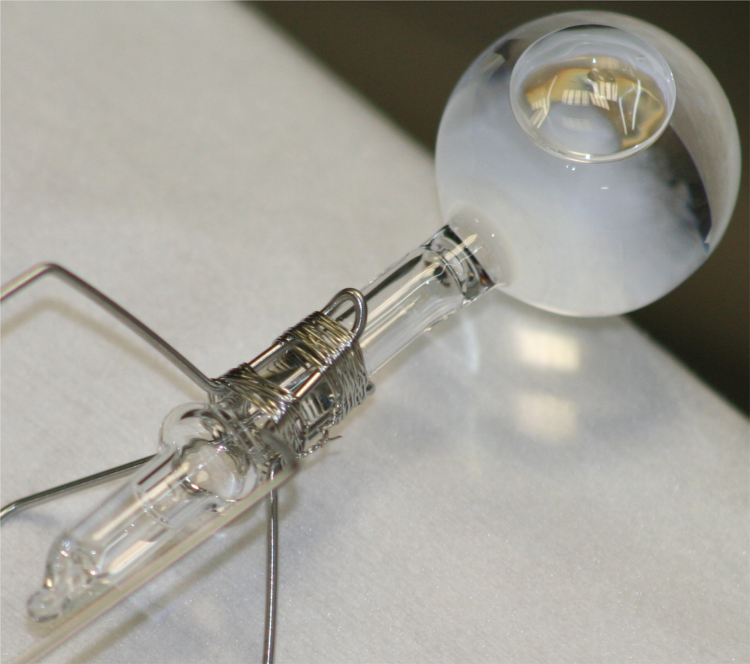}
\caption{The $^{203}$Hg $\gamma$ source that was deployed in June, 2009.  The spherical vial is made of quartz, whereas the neck is a graded transition from quartz to Pyrex.  The rounded bulge in the neck is a safety feature to prevent the source from slipping out.  Source retention is provided via the two independent wrappings with thin-gauge stainless steel wire.}
\label{fig:calib_source}
 \end{figure}

With the exceptions of the neutron and laser sources, the sources were
dissolved in either Borexino scintillator or water and sealed
within 1'' diameter quartz vials with a total volume of
$\sim$6\,ml. The neck of the vial had a graded transition 
to Pyrex glass for sealing purposes.  
After loading the vial with a radioisotope, the vial was frozen in liquid nitrogen, evacuated and flame-sealed.  
For instance, a close-up photo of the $^{203}$Hg source after being attached
to the source coupler is shown in Figure \ref{fig:calib_source}.

\subsubsection{$^{222}$Rn and $^{14}$C Sources}
\label{sec:rn222-c14-sources}

The vast majority of the calibration points were obtained with a compound 
\cfour-\radon\ source 
which provides $\alpha$, $\beta$, and $\gamma$ radiation across a 
large energy region. 
In this source, $^{14}$C and $^{222}$Rn were simultaneously
present in the scintillator.

Loading the vial with radon is accomplished via a 
calibrated flow-thru $^{222}$Rn source typically used 
for calibration of radon detectors. 
A source vial is evacuated and placed in a liquid nitrogen bath 
while a flow of radon-loaded nitrogen gas is established through the vial. 
Radon, whose melting point is at $202$\,K, freezes out onto the surface 
of the vial maintained at $77$\,K, while the carrier 
gas continues unaffected. 
After an appropriate build-up time has elapsed, the gas flow is 
ceased, the vial evacuated and warmed up. Finally, the 
vacuum in the vial is used to extract 
scintillator from a sparging flask until the vial is completely 
filled\footnote{The scintillator used to make the sources 
is taken from a sampling port on the Borexino purification system.  
The sample is maintained under an inert atmosphere at all times.}.

After this operation \cfour\, is loaded by using
a \cfour-toluene solution typically used for calibrating liquid 
scintillator detectors. The solution was pipetted into the source vial
filled with Rn-loaded scintillator. Since toluene
(methylbenzene) is chemically very similar to pseudocumene 
(trimethylbenzene) and homogeneously miscible, \cfour\ is
uniformly distributed throughout the vial.

The \cfour-\radon\ sources used in the calibrations were prepared at 
Virgina-Tech with 
an initial activity of $\sim$100\,Bq. 
They were deployed in more than 200 positions within the scintillator 
(see Section \ref{sec:int_source-locations}).

$^{14}$C is a $\beta$ emitter with an end-point energy of 156\,keV suitable for low energy studies.
The $^{222}$Rn  chain consists of three $\alpha$ emitters, namely $^{222}$Rn, $^{218}$Po and $^{214}$Po, with energies of
5.5\,MeV, 6.0\,MeV and 7.4\,MeV, respectively. Moreover, it has two 
$\beta$/$\gamma$ emitters $^{214}$Pb and $^{214}$Bi which have a $Q$ value of
1.0\,MeV and 3.2\,MeV, respectively.

\begin{figure}[htbp]
\begin{center}
\includegraphics[scale=0.6,angle=90]{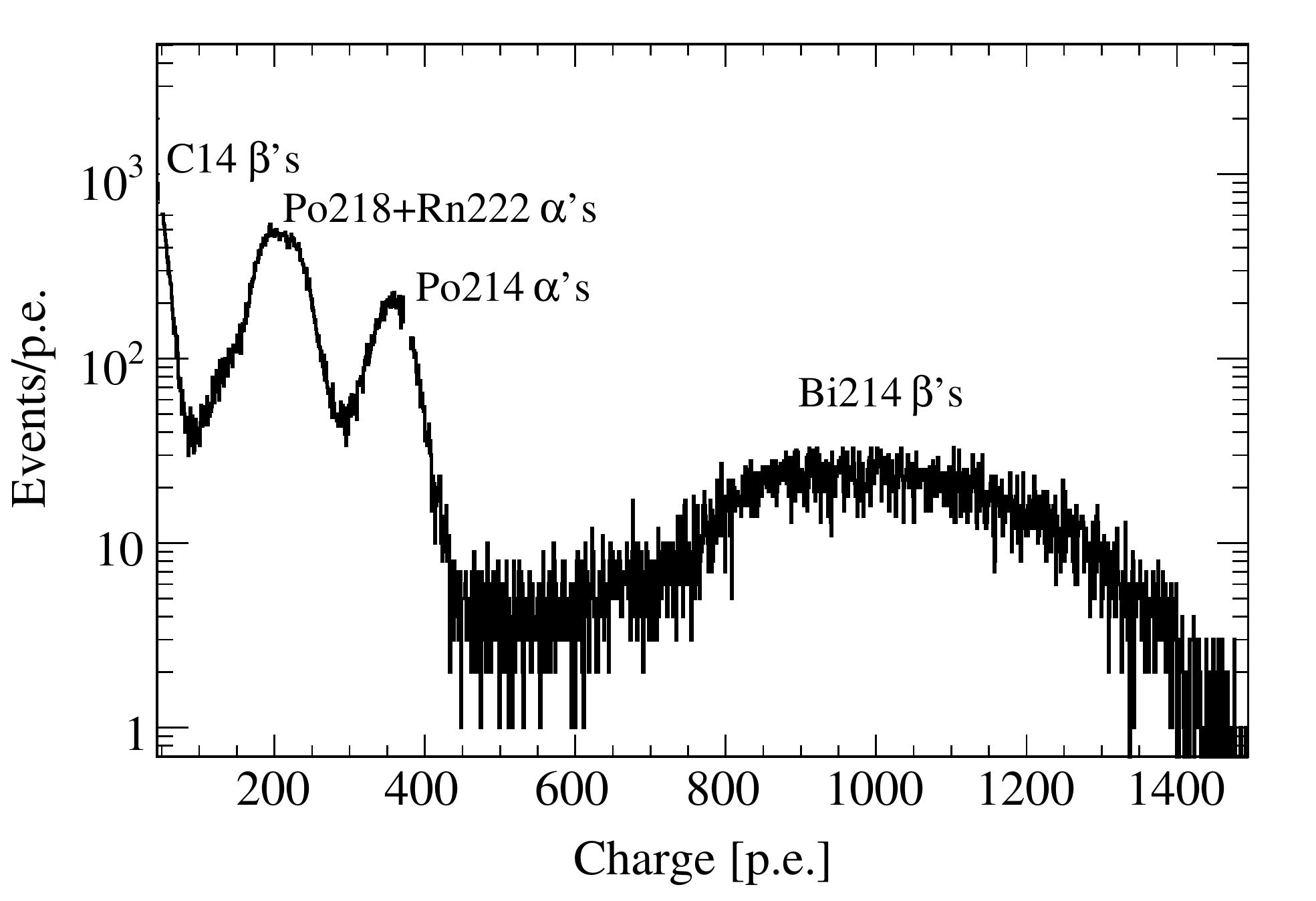}
\includegraphics[scale=0.6,angle=90]{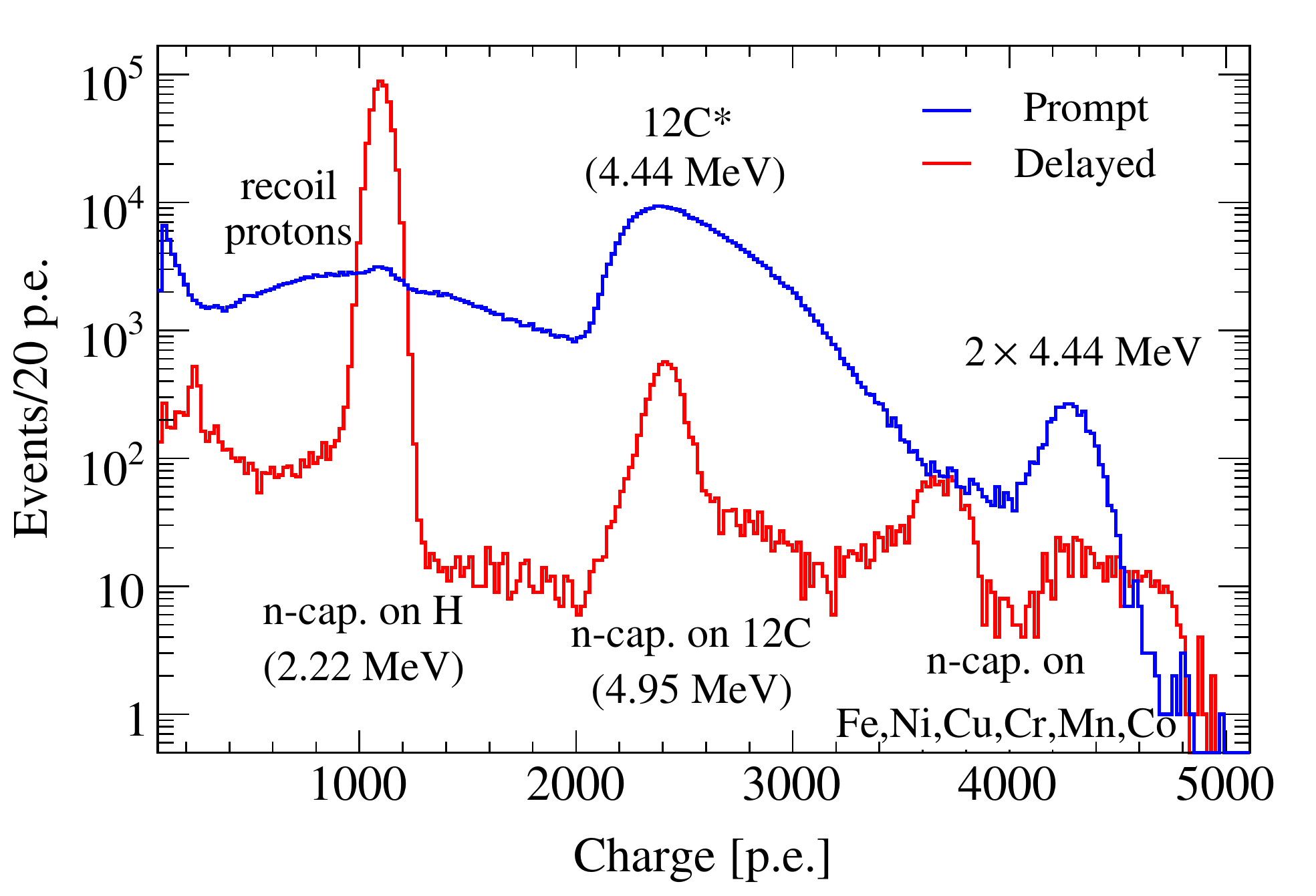}
\caption{\label{fig:srcspc} Energy spectrum (variable {\it p.e}) 
of the \cfour-\radon\ source (upper plot) and  $^{241}$Am$^9$Be source (lower plot).
 The $^{241}$Am$^9$Be spectrum is subdivided in a spectrum from neutron-induced prompt (blue) 
and delayed signals (red).}
\end{center}
\end{figure}

Due to the quenching phenomenon, $\alpha$ particles produce less 
scintillator light than $\beta$ particles (quenching factor $\sim$10 or more)
and therefore their equivalent energy in the detector is below 1\,MeV.
Therefore, the compound \cfour-\radon\ source provided calibration points in 
the energy region between 0 and 3.2\,MeV
which is relevant for the main solar neutrino analyses, i.e. $^7$Be, $pep$, and
CNO neutrino rate measurements. 
The upper plot in Figure~\ref{fig:srcspc} shows the spectrum of the 
\cfour-\radon\  source.

\subsubsection{$\gamma$ Sources}
\label{sec:gamma_sources}

In order to use $\gamma$ sources for energy calibration in Borexino, the sources had to be
mono-energetic and the scintillation light induced by associated $\alpha$ or $\beta$ radiation had to be suppressed.
This was achieved by depositing the radioisotope of interest in a non-scintillating 
medium within a source vial that absorbs the $\alpha$ and $\beta$ particles.

The isotopes of all $\gamma$ sources in Table \ref{tab:sources} were commercially obtained as salts dissolved in 
aqueous acid solutions.  
Thus, a pure $\gamma$ source was  obtained by pipetting the appropriate amount 
of isotope into a vial and filling the rest with deionized water. Due to EU radiation safety rules, the specific activity of the 
commercially-obtained solutions was limited to 1\,Bq/ml.

The activity of each of the $\gamma$ sources employed in the calibrations
was approximately 2\,Bq. For what concerns the $^{85}$Sr source,
the activity was precisely measured by means of $\gamma$ ray spectroscopy to
provide a reference value for trigger efficiency studies 
(see Section \ref{sec:trigger}).

\subsubsection{$^{241}$Am$^{9}$Be Neutron Source}
\label{sec:ambe_source}

\begin{figure}[htb]
\centering
\includegraphics[width=0.4\textwidth]{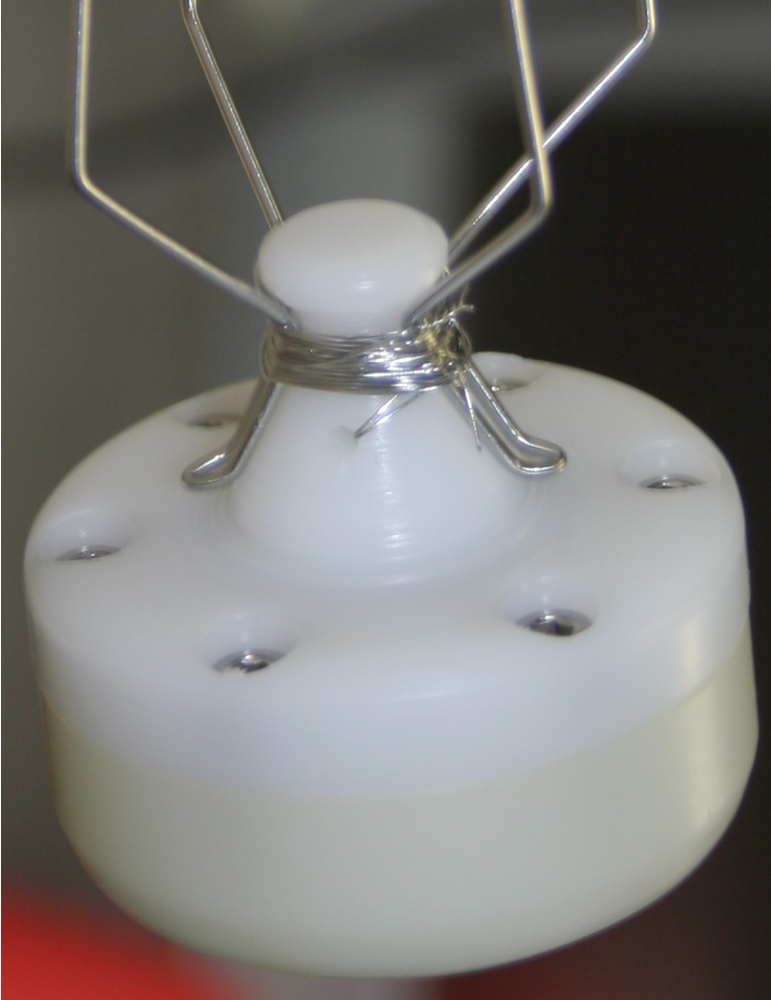}
\caption{Photo of the encapsulation in which the Borexino $^{241}$Am$^{9}$Be neutron source was placed. Note that
the neck contains the same mounting features as the regular source vials.}
\label{fig:neutron_source}
\end{figure}

For $^8$B neutrino and geo-neutrino analyses \cite{borex_b8, borex_geonu}, for studies of solar and other unknwon anti-neutrino fluxes \cite{borex_antinu}, and search for solar axions \cite{borex_axion} it is essential to have energy calibration points up to 10\,MeV. In Borexino this was accomplished by a $\sim$10\,Bq $^{241}$Am$^{9}$Be neutron source that was inserted into the detector
during the second and third internal calibration campaigns.  As shown in Figure \ref{fig:neutron_source}, the source was contained in 
a 3\,mm lead capsule embedded in a delrin holder to shield the 60\,keV $x$ rays (activity $\sim$180\,kBq) produced by the source itself.
 
Neutrons are produced in two main reactions, $^{9}$Be($\alpha$,n)$^{12}$C$_{gs}$ 
and $^{9}$Be($\alpha$,n)$^{12}$C$^*$ (4.44\,MeV) with energies up to 11\,MeV and 6.5\,MeV,
respectively. The second reaction also produces one or two $\gamma$ rays with a total
energy of 4.44\,MeV from the $^{12}$C$^*$ de-excitation.
These $\gamma$ rays, together with the recoil protons from neutron
scattering in the medium, are responsible for a prompt scintillator signal. 

Afterwards, neutrons thermalize in the hydrogen-rich organic liquid  
and are captured either on protons or carbon nuclei 
in the scintillator emitting characteristic 2.22\,MeV and 4.95\,MeV $\gamma$ rays, respectively.
These characteristic $\gamma$ rays produce a
 delayed signal in the scintillator according to the neutron capture time of $\sim$254\,$\mu$s in pseudocumene
 \cite{neutron-muon-paper}.
In addition, neutrons are captured on iron, nickel and chromium nuclei of the stainless steel source insertion arm 
resulting in the emission of $\gamma$ rays with energies up to 9.3\,MeV. 
The lower plot in Figure~\ref{fig:srcspc} shows the energy spectrum of both the prompt and delayed signals produced by the 
$^{241}$Am$^{9}$Be source.

\subsubsection{Laser Source}
\label{sec:laser-source}

During the third internal off-axis calibration campaign and during a 1-day run 
in September 2009 a 
394\,nm laser was used to provide pulsed light to a diffuser ball
located at the center of the detector. The diffuser was similar to the one used 
by the source location system, but was designed on purpose to provide
better light uniformity.
The primary goal of the laser source calibration was to check 
the PMT time synchronization independently
from the main equalization system. 
The standard PMT equalization is performed routinely by means of a multiplexed
optical fiber system which delivers the laser light
directly and simultaneously on each photocathode (more details of this system can be found in
\cite{PMT_timing}). 
During the laser source calibration runs, the laser and the detector were
synchronously triggered at 50\,Hz at several different laser intensities. 
 

\subsection{Internal Calibration Campaigns}
\label{sec:int_cam}

\subsubsection{Goals}
\label{sec:int_goals}

The internal source deployment and location systems described in the previous section
were used in four calibration campaigns between 2008 and 2009: one on-axis campaign with modified hardware in October 2008 (I), 
where the sources were deployed only on the vertical z-axis through the detector center, 
and three off-axis campaigns in January 2009 (II), June 2009 (III), and July 2009
(IV). 
The major goals of these calibration campaigns were:
\begin{itemize}
\item Determination of the energy scale;
\item Study of the uniformity of the detector response;
\item Testing the position reconstruction algorithms and study of FV systematics at different energies and for
different FV cuts;
\item Validation of the Borexino Monte Carlo code and of the analytical models describing the detector response
that were used during signal extraction; 
\item Study of the detector trigger efficiency.
\end{itemize}

\subsubsection{Source Locations}
\label{sec:int_source-locations}

\begin{figure}[htb]
\centering
\includegraphics[width=0.8\textwidth]{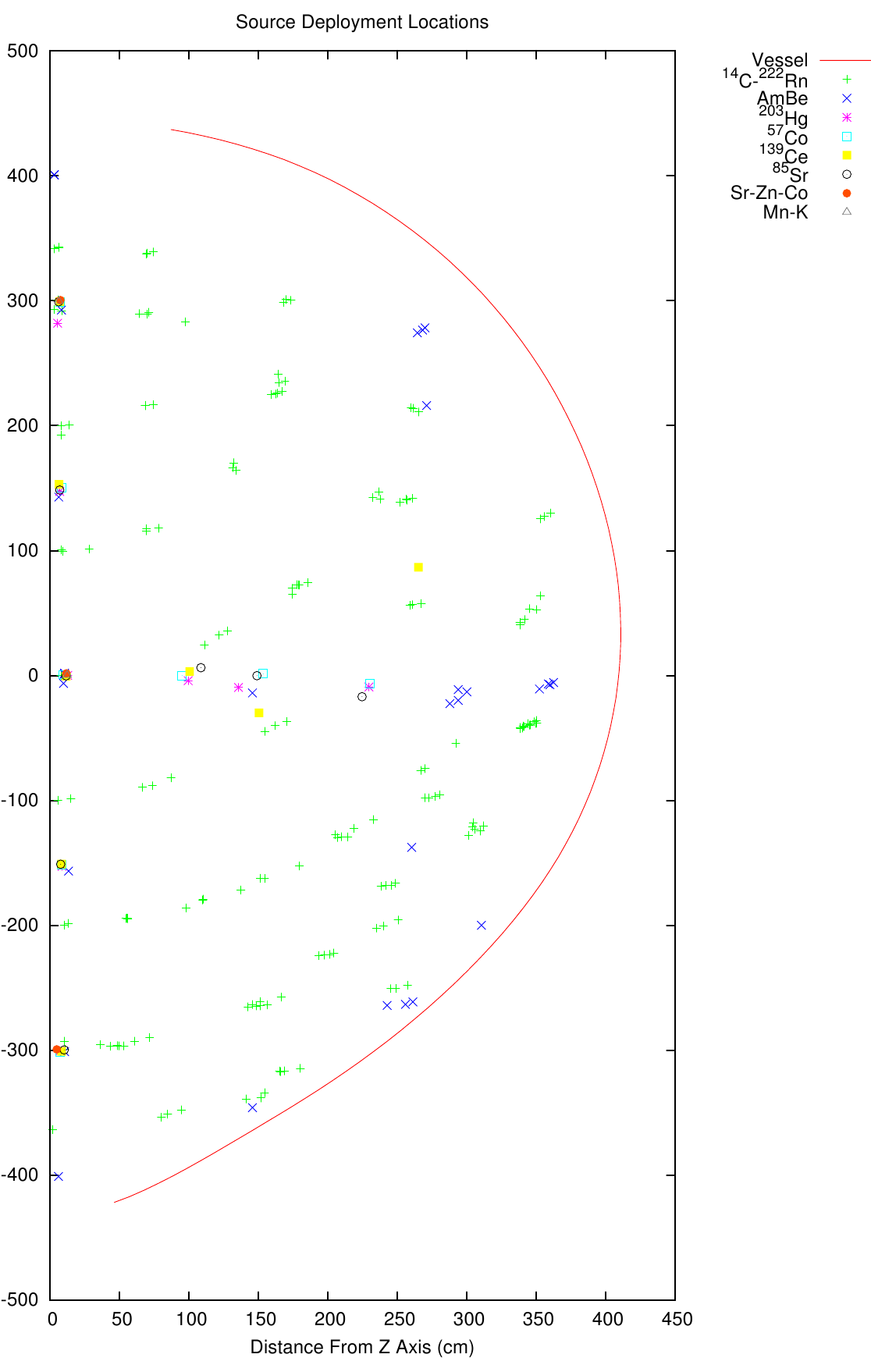}
\caption{A map of the locations where sources were deployed during
the four internal calibration campaigns. The horizontal axis corresponds to the distance from
the vertical $z$-axis. In total, the sources were deployed in over 200 locations. 
Several points illustrated in the picture were also repeated at different $\phi$ coordinates. 
The red line indicates the shape of the inner vessel on January 22, 2010.}
\label{fig:deploy_loc}
\end{figure}

In order to perform position studies, some of the sources were 
deployed in different locations throughout the
scintillator volume: overall, during the four calibration campaigns 
twelve different sources were deployed 
in 295 locations.    
In order to study the energy scale, the eight monochromatic 
$\gamma$ sources (see Table~\ref{tab:sources}) were
deployed at the center of the detector and in few positions along the z-axis.
Some of the sources were also deployed in other regions of the
scintillator volume (mainly at larger distance from the center) in 
order to probe the uniformity of the trigger efficiency
of the detector. 
A fine-grained mapping of the detector response
was performed using using the \cfour-\radon\ source deployed
in more than 200 positions (see Figure ~\ref{fig:deploy_loc}).

Since the shape of the inner vessel changed significantly 
during the first three years of data-collection, it was important to ensure that these 
instabilities did not affect the optics of the detector and thus 
the position reconstruction performance. 
As a consequence, \cfour-\radon\, sources were deployed in approximately 
ten standard locations within the scintillator during each calibration campaign. 
No significant time variation of the event position reconstruction 
in these standard positions was observed.

Most off-axis points scanned with the \cfour-\radon\, source were located 
at three equidistant $\phi$ values ($\Delta\phi$=120$^\circ$) for a given ($R$, $\vartheta$) pair. 
However, a more fine-grained mapping ($\Delta\phi$=30$^\circ$) was performed
on the equatorial plane to study possible biases present 
in the $\phi$ coordinate.
Since Borexino approximately employs a spherical FV cut at $R$$<$3.021\,m for its $^7$Be neutrino analysis,
the \cfour-\radon\, source was used to map the volume in intervals of  
$\Delta\vartheta$=15$^\circ$ and $\Delta\phi$=90$^\circ$.
These data were also used to map the precise radius of the FV as a 
function of the event energy. This was fundamental for the reduction of the systematic
uncertainty on the FV (compare with Section~\ref{sec:PRC}).

In order to perform position studies at higher energies, the
custom-made $^{241}$Am$^{9}$Be was deployed in approximately 30 positions within the scintillator.

Finally, during the last calibration campaign a pulsed laser source was employed at the center of the detector. Its main purpose was
to perform an independent check of the PMT time equalization.

\subsubsection{Procedures}
\label{sec:int-calibr-procedure}

Each of the off-axis calibration campaigns lasted approximately two weeks, 
and involved the deployment of about five sources each time. A detailed description about it can be found in \cite{SH_Dissertation}.

The entire process of the source removal and re-insertion 
takes at least two hours and for that reason all 
the points that required four rods below the hinge were 
performed first, followed by the three-rod ones, and so on.
The source insertion procedure begins with the attachment of a source to the 
source coupler and sealing the 6-way cross.  To establish an inert
atmosphere in the cross, it is evacuated and then backfilled with 
LAKN$_2$ at least eight times.  Once the
required pressure and cleanliness are established, the insertion begins after opening the gate valve.
When the source is finally deployed at a given position within the inner vessel, a 
photo is taken to confirm its position. This process is repeated for each rod inserted or retracted.  
Once the hinge has reached its desired position along the z-axis, 
the rod depth is locked with a special clamp and the $\phi$ angle is set. 
Next, the program calculates the amount of tether that needs to be 
extracted in order to establish the $\vartheta$ angle.

The source position is checked with a photo before starting  data-acquisition. 
After the completion of a DAQ run,  the location of the source is verified one more time before moving to a new position. 
The extraction procedure corresponds, in principle, to a reverse order of the insertion steps.
However, special precautions are required to assure that the source passes the gate valve level. For this purpose, a special
sweep-arm situated above the closure-path of the gate valve is used, as shown in Figure \ref{fig:sweep-arm}.

\subsubsection{System Performance}
\label{sec:int_sys_performance}

Overall, the four internal calibration campaigns lasted for 
54 days with a duty cycle of 65\%.

The initial part of the first  internal calibration was devoted 
to commissioning, including improvements on the process control system in terms of 
safety and stability. Moreover, the diffuser and illumination method was found to be inefficient 
and a more effective diffuser and laser-coupling scheme was developed for 
the following calibration campaigns.
During the first calibration campaign, failure of one of the CCD cameras 
occurred. However, since the source location system was intentionally redundant, the loss of a 
single camera did not reduce the overall precision of the source position determination.

\paragraph{Preservation of System Cleanliness}
\label{sec:SC}

Even though the Borexino  internal calibration system was 
designed and built following very strict requirements on radiopurity 
(see Section \ref{sec:int_clean}),
calibration activities always carry a non-zero risk of contamination.
For this reason,  internal calibrations were not performed at the beginning of 
Borexino data-acquisition, but after more than one year, 
when a significant sample of good data had been collected.

\begin{figure}[htb]
\centering
\includegraphics[scale=0.6,angle=90]{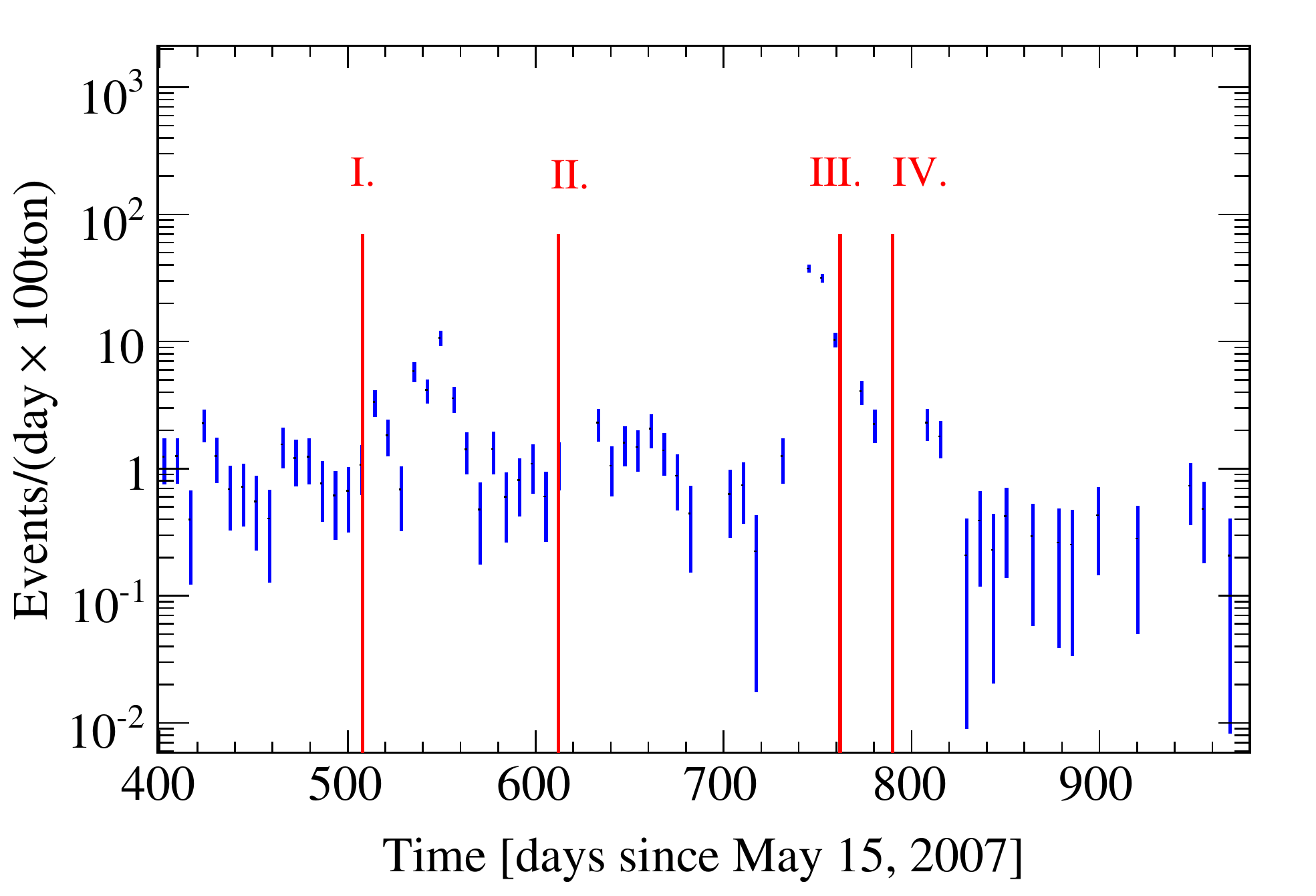}
\caption{Rate of $^{214}$Bi$^{214}$Po coincidences in the FV as a function of time. 
The peaks corresponds to Radon spikes during operations. The red lines mark the time when the four 
 internal calibration campaigns were carried out.}
\label{fig:csu} 
 \end{figure}

\begin{figure}[htb]  
\centering
\includegraphics[scale=0.6,angle=90]{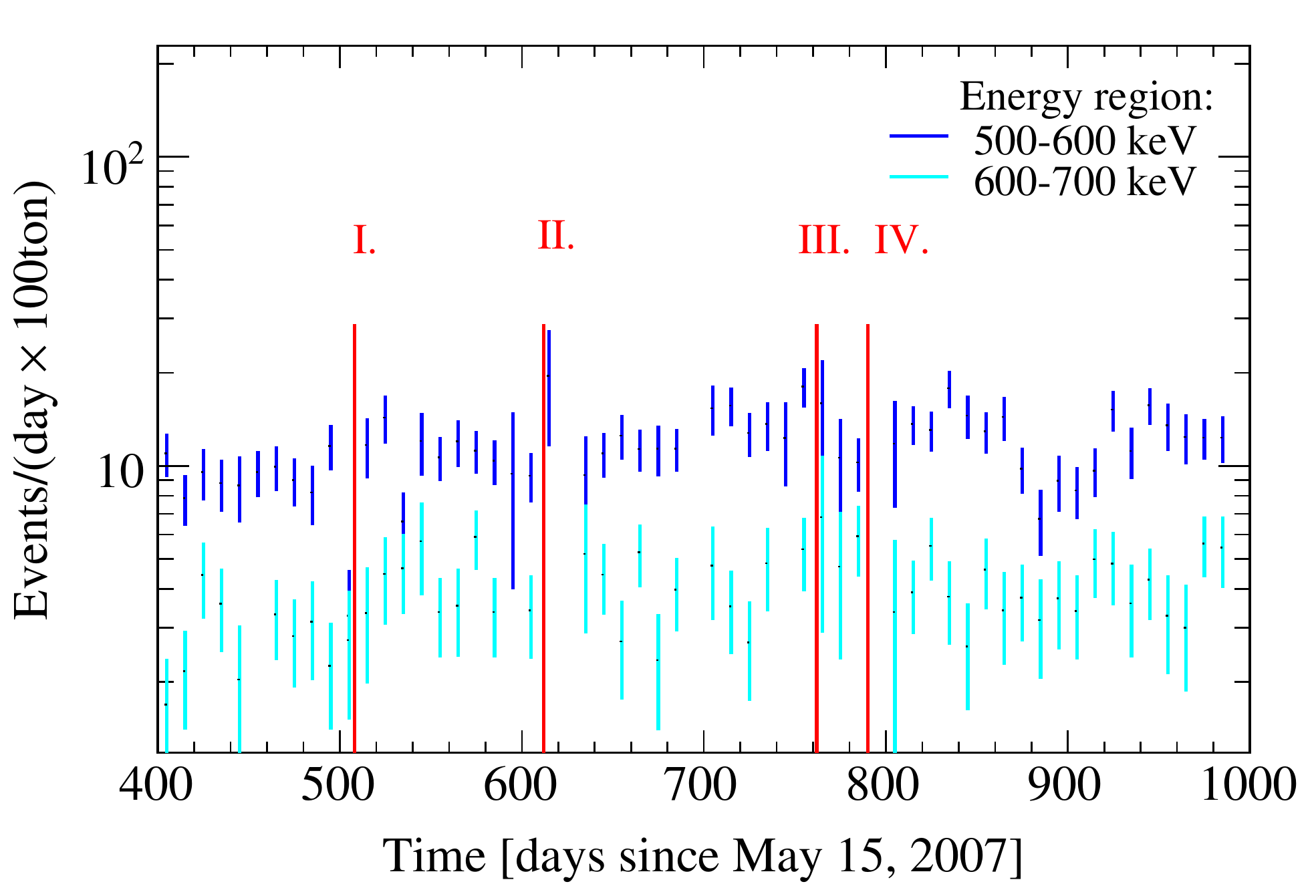}
\caption{Count rate in the energy region between (500-600)\,keV and (600-700)\,keV  
as a function of time. The start of data-collection in Borexino was May 15, 2007. The red lines mark the time period when the four
 internal calibration
campaigns were carried out.}
\label{fig:val}
\end{figure}

The full-scale  internal calibration program
left no detectable evidence of impurities.
This can  be demonstrated  by analyzing data acquired before 
and after each calibration campaign.
In particular, the $^{238}$U concentration in the scintillator 
can be measured by tagging the $^{214}$Bi$^{214}$Po fast decay coincidences
($\tau$=236.6\,$\mu$s) that belong to the $^{238}$U chain. 
Figure \ref{fig:csu} shows the coincidence rate (Unit: events/(day$\times$100\,ton)) 
as a function of time, after
applying the same fiducial volume cut used in the $^7$Be neutrino analysis 
($R$$<$3.021\,m; $|z|$$<$1.67\,m). 
The red lines mark the beginning of each calibration campaign. 
No evidence for an increase in the $^{238}$U contamination due to 
calibrations is found. 
Several transient spikes in the coincidence rate can be observed. They are due to the insertion of Radon during 
operations (including calibrations) performed on the detector. 
Since Radon decays fast 
($\tau$=5.48\,d) this does not affect the long-term scintillator radiopurity
\footnote{Build-up of $^{210}$Pb (and therefore $^{210}$Bi) 
due to the transient Radon spikes is
expected to be negligible ($<$ 2 counts/day/100tons) given the small amount of
Radon inserted. Indeed the
count rate before and after calibrations in the energy regions which would be
affected by $^{210}$Bi has remained constant as shown in
Figure~\ref{fig:val}.
}.
From the coincidence analysis, the $^{238}$U concentration 
before and after the campaign of calibrations is estimated to be (5.0$\pm$0.9)$\times$10$^{-18}$\,g/g and 
(3.2$\pm$0.7)$\times$10$^{-18}$\,g/g, respectively.

Similarly, the $^{232}$Th concentration can be investigated by looking at the fast $^{212}$Bi$^{212}$Po coincidences. As before, this analysis shows
no evidence for an increased $^{232}$Th contamination. Indeed, 
the overall contamination in $^{232}$Th before and after
all four  internal calibrations was found to be (3.0$\pm$1.0)$\times$10$^{-18}$\,g/g and (5.0$\pm$1.5)$\times$10$^{-18}$\,g/g, respectively.

One of the most dangerous background for the $^{7}$Be neutrino analysis is $^{85}$Kr, a $\beta$ emitter with 
a $Q$ value of 687\,keV. This isotope can also decay with a very small branching ratio (0.43\%) into a metastable state of $^{85}$Rb, 
which de-excites emitting a characteristic 514\,keV $\gamma$ ray after $\sim$1 $\mu$s. This rare coincidence tag can 
be used to evaluate the $^{85}$Kr concentration in the scintillator.
Since the number of such coincidences is minimal, i.e. a few events within a year of data-acquisition, this method cannot be used to evaluate the impact 
of each single calibration campaigns. However, the coincidence analysis was able to show that the $^{85}$Kr contamination before and after
the four calibration campaigns remained constant within its total uncertainty budget.

Concerning other contaminants affecting the solar neutrino analysis, a 
study was performed on the stability of the count rate in two energy
regions, namely (500-600)\,keV and (600-700)\,keV. 
These energy regions are relevant for the $^7$Be as well as
for the {\it pep} and CNO neutrino analyses. Approximately 55\% of the count
rate in the first window is due to $^7$Be neutrinos, while the remaining 45\% is due
to $^{85}$Kr and $^{210}$Bi, a $\beta$ emitter with an end-point energy of 1.16\,MeV.
In the second energy window  around 40\% of the count rate is due to $^7$Be,
CNO and {\it pep} solar neutrinos, while the remaining counts are due to $^{210}$Bi decays. 
Figure~\ref{fig:val} shows the count rate in the FV defined for the $^7$Be neutrino analysis as a
function of time. The rates for both energy windows are shown in dark and light blue, respectively. 
The plot shows no evidence for an increase of the counting rates due to the calibration campaigns.

In summary, the Borexino internal source deployment system has functioned as designed,
by deploying a number of different sources at more than 200 locations within the IV,
without leaving any detectable radioactivity in the scintillator.

\section{External Source Calibration}
\label{sec:ext_calibr_system}

In addition to the invasive internal calibration system 
described in the previous section,
the Borexino detector is equipped with a second non-invasive calibration 
system, used to deploy $\gamma$ sources in
the outer buffer region.
The goal of this system is to study the 
spectral shape and the radial dependence of the external background 
events caused
by radioactivity in the outer detector components, such as 
PMTs, light concentrators, stainless
steel sphere and so on.
The most important external background in the Borexino analysis is the
2.615\,MeV $\gamma$ rays produced in decays of $^{208}$Tl. 
Since these $\gamma$ rays cannot be easily tagged and discriminated,
they can severely affect the outcome of physics results (see, for instance, \cite{borex_b8, borex_pep}). 

Section \ref{sec:ext_calibr_system_hardware} describes the hardware 
of the external calibration system, while
section \ref{sec:$^{228}$Th_source} discusses the custom-made 
$^{228}$Th source used for the two  external calibration campaigns.
Section \ref{sec:ext_calibr_camp} gives details on the
procedures adopted during the calibration operations.
The  external calibration campaigns provided information not 
only on the external background, but also on global detector properties. 
This is shown in Section \ref{CR}.

\subsection{Hardware}
\label{sec:ext_calibr_system_hardware}

\begin{figure}[htb]
\centering
\includegraphics[scale=0.4]{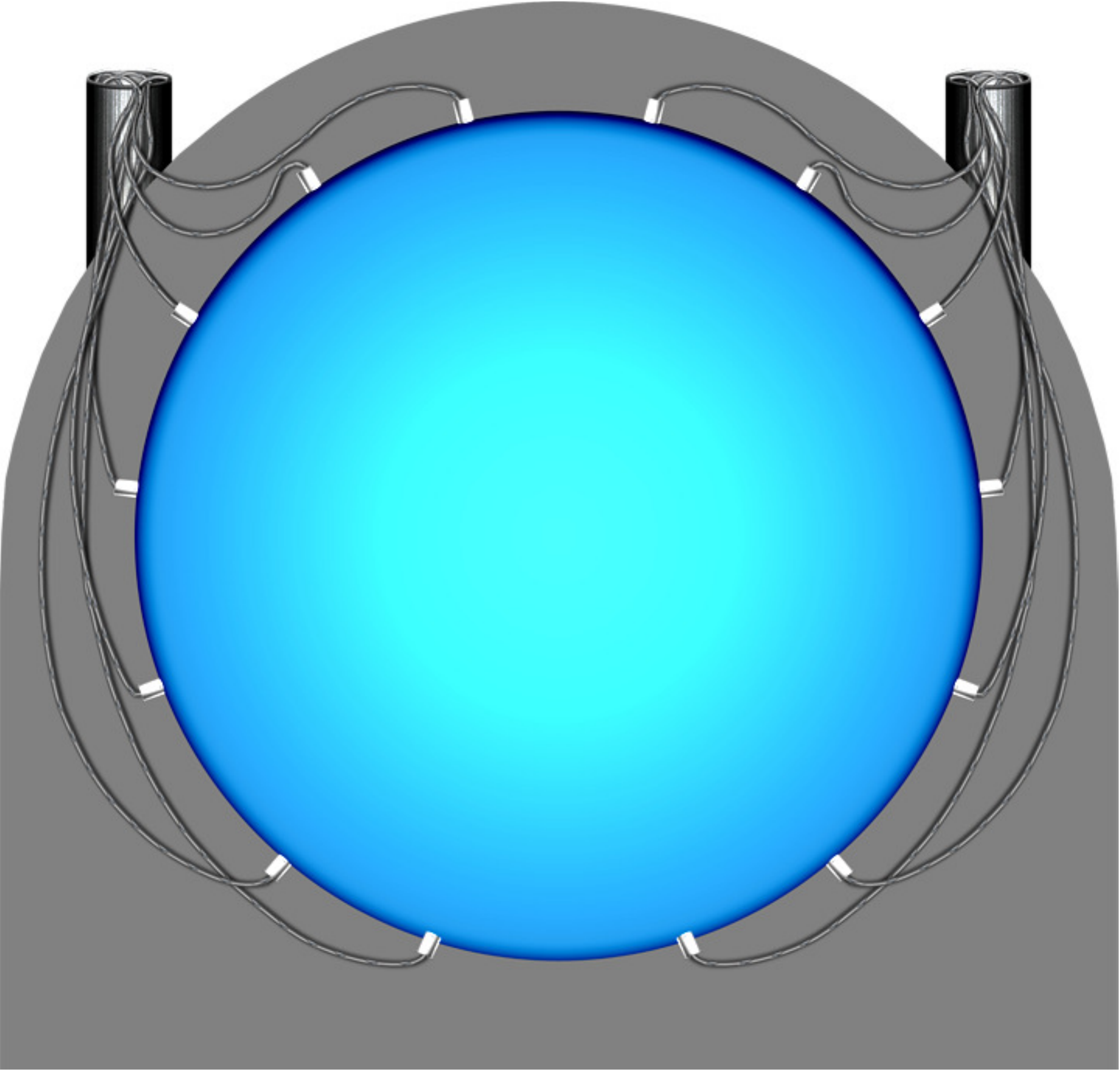}
\caption{The geometry of the external source insertion system. 
Polyethylene tubes (grey) connect the top of
the detector to fourteen reentrant tubes mounted on the SSS (blue) 
and terminate at the same depth as the
PMTs at a distance of 635\,cm from the center of the detector. 
A small $\gamma$ source can be placed inside of a
specially designed capsule attached to an electrician's fish tape 
and fed down through the tubing to the end
of a reentrant tube. More details about this calibration system are summarized in the text.}
\label{fig:ext_src_geom}
\end{figure}

The external calibration system consists of fourteen 61\,cm long reentrant tubes mounted at different
positions along a vertical plane on the stainless steel sphere. 
The reentrant tubes are connected with
flexible polyethylene tubing going through the water tank 
and ending in an organ pipe on the top platform
of the detector. Figure \ref{fig:ext_src_geom} illustrates 
the geometry of the system. It allows to
position the source at the level of the PMTs to a minimal 
distance of $\sim$635\,cm from the center
of the detector. The tubes are thick enough to withstand the buoyant 
forces to which they are exposed due to the presence of water
in the outer detector. Additionally, the inner surface 
of the tubes was made smooth in order to reduce friction
when a source is moved through it. 

\begin{figure}[htb]
\centering
\includegraphics[width=0.9\columnwidth]{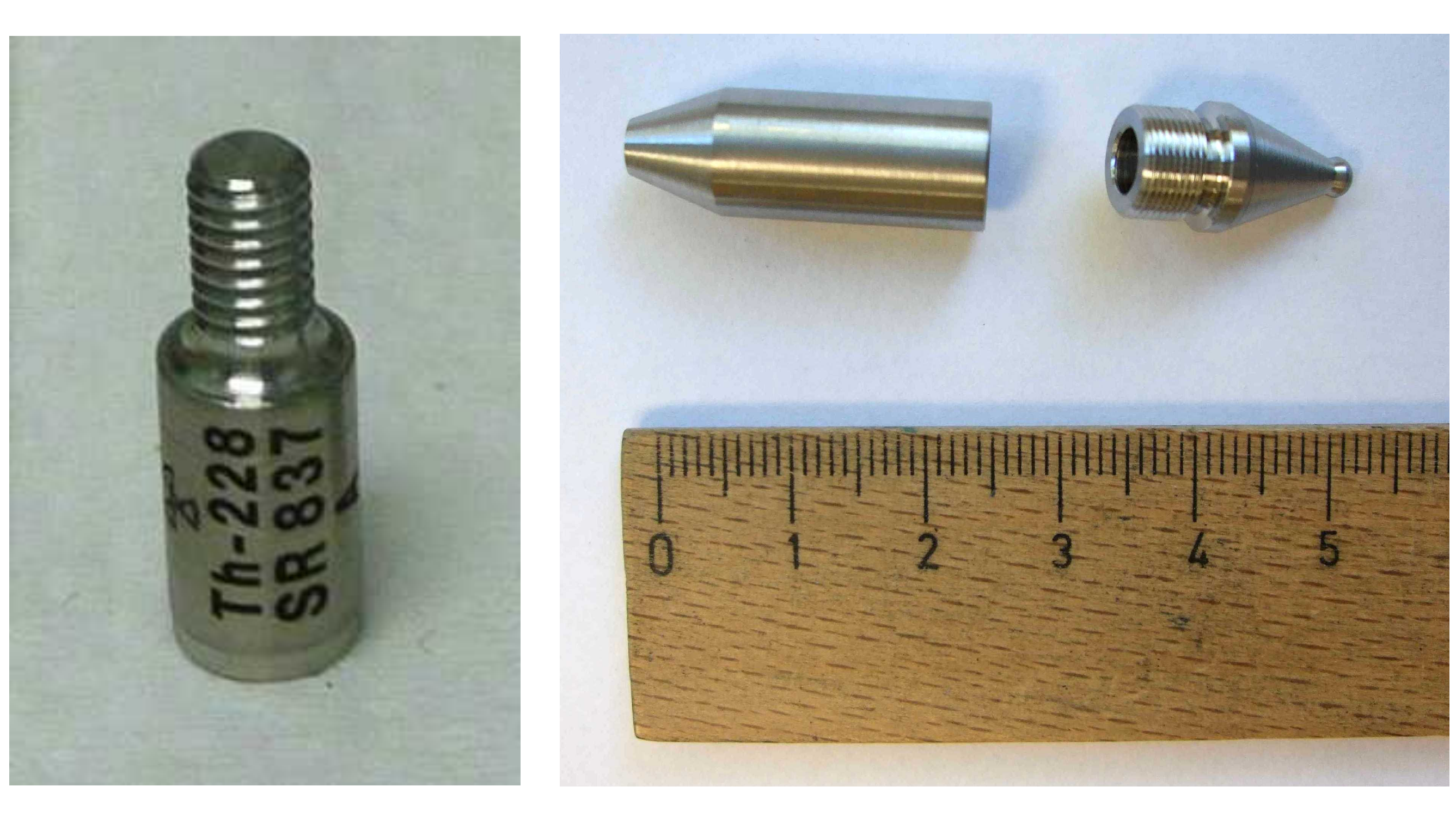}
\caption{Custom-made $^{228}$Th source (left) and capsule (right) for holding the source to be deployed by 
the external source system. The left edge of the capsule is silver-soldered onto a metal electrician's fish tape 
and then pushed through one of fourteen tubes until reaching the buffer region in the inner detector. Units of the capsule
are given in centimeters.}
\label{fig:ext_src_capsule}
\end{figure}

The external source deployment begins with placing the source 
inside a specially designed capsule shown in Figure \ref{fig:ext_src_capsule}. 
The capsule is made of stainless steel and is silver-soldered
to a metal electrician's fish tape. The diameter of the capsule 
is 9\,mm. The capsule is wrapped in electrical 
tape and the fish tape is then fed into the
polyethylene tube until it reaches the end of the reentrant tube 
on the sphere. A completed insertion is verified
with a closed circuit between the fish tape and 
the flange on the organ pipe, i.e. the circuit
is established when the end of the metallic capsule touches the 
end of the reentrant tube. Based on this reference
point it is also possible to move back the source within the 
reentrant tube to different positions that are
known with an accuracy of $\pm$3\,cm.

\subsection{Custom-made 5.41\,MBq $^{228}$Th Source}
\label{sec:$^{228}$Th_source}

The radioactive source that was chosen for the  external calibrations of
Borexino is a $^{228}$Th source: $^{228}$Th is a relatively long-lived nuclide ($\tau$=2.76\,y) and 
 $^{208}$Tl is one of its daughter nuclides: 
 the emission probability of the 2.615\,MeV $\gamma$ energy line is 35.6\%.
 
In order to collect enough statistics in a reasonable time frame 
a $^{228}$Th source with an activity of several
MBq was needed. Commercially available sources of such activities 
typically use ceramic matrix materials to
incorporate the radionuclides. 
Depending on the chemical composition, the low-Z nuclides in the ceramics 
induce the emission of unwanted
neutrons due to ($\alpha$,n) reactions. Even though a neutron source 
strength of the order of 100\,s$^{-1}$ is
not problematic for Borexino, strong limitations for the usage 
of neutron emitting sources are imposed by
the LNGS underground laboratories. The reason is that neighboring 
experiments searching e.g. for
dark matter might be perturbed by these neutrons. 
Due to this fact, the production of a custom-made
$^{228}$Th source with reduced neutron source strength became necessary. 
It was achieved by using a gold
foil in which the thorium was embedded. Gold has a high ($\alpha$,n) 
energy threshold of 9.94\,MeV, while
the mean energy of $\alpha$ particles from the $^{228}$Th chain is 6.5\,MeV.
A detailed description of the source
production and characterization is given in \cite{borex_thorium_source}. 
In summary, the final activity of the
source was (5.41$\pm$0.30)\,MBq, while the neutron source strength
was measured to be (6.59$\pm$0.85)\,Bq
(Reference date: March 1, 2010).

\subsection{External calibration campaigns}
\label{sec:ext_calibr_camp}

\subsubsection{Goals}
\label{ext_goals}

The external calibration system and the custom-made $^{228}$Th source 
described in the previous sections were used in two external calibration campaigns. 
The main goals of these calibration measurements were:

\begin{itemize}
\item Determination of the spectral shape of the external 2.615\,MeV $\gamma$ rays for different fiducial volumes;
\item Study of asymmetries in the energy response of the detector 
for events more distant from the detector center;
\item Determination of the radial distribution of the external $\gamma$ events;
\item Precise determination of the inner vessel shape;
\item Validation of the Borexino Monte Carlo code.
\end{itemize}

In addition, other large-scale scintillator and inner detector (ID) properties could be determined. Some of the obtained results are
presented in Section \ref{sec:ext-bkgr}.

\subsubsection{Procedures and system performance}
\label{sec:procedure_ext_calibr}

The first external calibration campaign was performed in July 2010
for a period of 9.0 days with a duty cycle of 95\%.
At that time the activity of the source was $\sim$4.8\,MBq. 
Data were collected in three different positions, two lying
in the upper and one in the lower hemisphere of the detector. 
In one case the source was at a distance of $\sim$635\,cm  from
the center of the detector, in the other two cases at $\sim$685\,cm, i.e.
close to the stainless steel sphere. During the calibration campaign,
the Borexino trigger threshold was increased from $\sim$50\,keV 
to $\sim$200\,keV in order to guarantee a stable DAQ. Details are given in \cite{WM_Dissertation}.
Moreover, the most exposed PMTs close to the source positions 
were tested showing no anomaly during the whole measurement
despite the high trigger rate.

In order to study the properties of the external background in Borexino 
in more detail, a second  external calibration
campaign was carried out in November-December 2011. It lasted for 27.9 days
and the achieved duty cycle was 80\%. 
During this calibration campaign, a relatively low trigger threshold of $\sim$120\,keV
was adopted in order to keep the detector supernovae-live for different neutrino and antineutrino flavors \cite{SN-paper}.

At the time of the second calibration the activity of the $^{228}$Th source was still $\sim$2.9\,MBq. 
Data were acquired for ten different positions distributed from the top
to the bottom of the detector. In all cases the distance of the source from the detector center was $\sim$685\,cm.
Several positions were measured twice for stability tests at the beginning and at the end of the
calibration campaign. No change in the detector performance was observed.

\section{Calibration Results}
\label{CR}

In this Section we show that the calibration 
systems described in Section \ref{sec:int} and \ref{sec:ext_calibr_system}  
have worked properly 
and have significantly contributed to the success of the Borexino experiment. 
A selection of several important issues which have been thoroughly studied
using calibration data is presented. 
Sections \ref{sec:ER} and \ref{sec:results_position-reco-fv}  discuss 
some of the systematics studies performed using data collected during 
the internal calibration campaigns.
Section \ref{sec:ER} briefly reviews the role that calibrations have played on the energy scale determination, while
 Section \ref{sec:results_position-reco-fv} is focused on the tests and  
 validation of the event position reconstruction algorithm. 
Section \ref{sec:trigger} is focused on the evaluation of the trigger
efficiency as a function of the reconstructed event position for energies below
$\sim$0.5\,MeV.
Section 5.4 summarizes an interesting application of both 
the CCD camera location system and the
external calibration system for monitoring the inner vessel shape.
Finally, Section ~\ref{sec:ext-bkgr}, discusses results obtained from the external calibration campaigns such 
as the energy spectrum and radial distribution of external $\gamma$ rays within the buffer and scintillator. Furthermore,
several global detector properties
derived from the  external calibration data will be presented.

\subsection{Energy Reconstruction}
\label{sec:ER}

\begin{figure}[htbp]
\begin{center}
\includegraphics[scale=0.6,angle=90]{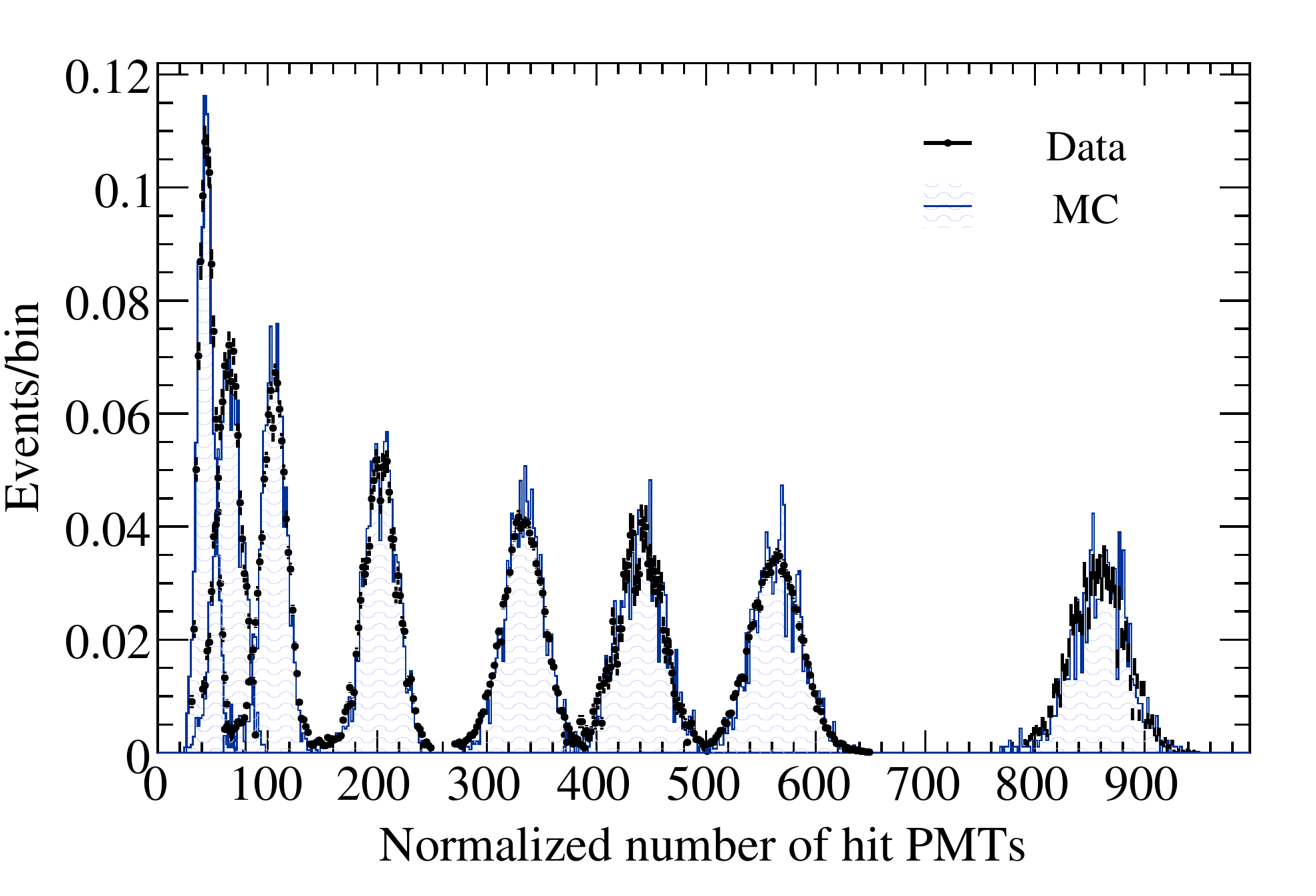}
\end{center}
\caption{\label{fig:calib_E} Energy spectra of $\gamma$ lines from eight different 
calibration sources expressed in terms of the normalized number of hit PMTs. 
The peaks from left to right belong to the $^{57}$Co, $^{139}$Ce, $^{203}$Hg, $^{85}$Sr, 
$^{54}$Mn, $^{65}$Zn, $^{40}$K, and $^{60}$Co source. The area of the single peaks are normalized to unity for a better comparison. 
The Monte Carlo simulated spectra are within 0.2\% agreement with the measured ones.}
\end{figure}

In principle, the energy deposited by a particle interacting in the 
Borexino scintillator is proportional 
to the number of photons collected by the PMTs.
An electron with kinetic energy of 1\,MeV produces approximately 500\,photoelectrons  
in the Borexino detector. 
We define here two energy estimators which will
be used in the following: {\it p.e.}, 
the total number of photoelectrons collected in the acquisition
gate and $N_{PMT}$, the number of triggered PMTs. 
The light production in an organic scintillator
is affected by non-linear effects such as light quenching \cite{Birks}. 
Furthermore, due to the large size of the detector,
 light collection is significantly perturbed
by photon propagation effects like absorption, re-emission, and scattering processes. In addition, geometrical
shadowing effects and an asymmetric distribution of operational PMTs have to
be taken into account. All these effects can contribute to a non-uniform energy response.
For these reasons, the determination of the energy scale of the Borexino 
detector is complex and requires Monte Carlo simulations
capable of reproducing the details of the emission and the propagation 
of photons within the scintillator.
The internal calibration campaigns were fundamental in this respect, 
allowing us to fine-tune several 
Monte Carlo input parameters like the light yield, the quenching factor k$_B$ according to the Birks 
model \cite{Birks}, and the scintillator and buffer attenuation lengths.
The energy scale determination and its uniformity
 will not be covered here. A comprehensive discussion, including all 
 the details concerning the Monte Carlo code and its optimization can be found in  
\cite{MC_paper}.
As an example, Figure~\ref{fig:calib_E} shows 
the energy spectra for the eight $\gamma$ sources described 
in Section~\ref{sec:sources}.
These sources provided eight calibration points
covering most of the energy region of interest for Borexino, i.e., 
from as low as 122\,keV ($^{57}$Co source)
up to 2.5\,MeV (sum of the two $\gamma$ rays from $^{60}$Co decays). 
This was essential to fine-tune the Monte Carlo code which should 
reproduce the
scintillator non-linear response over this relatively large range of energies.
The plot  shows the positions of the $\gamma$ energy lines 
 obtained from the calibration data and from Monte Carlo simulations:
the agreement between the two is excellent. 

\subsection{Position reconstruction and fiducial volume}
\label{sec:results_position-reco-fv}

\subsubsection{Position Reconstruction: Algorithm and Effective Index of Refraction}
\label{sec:PRP}
 
\begin{figure}[htbp]
\begin{center}
\includegraphics[scale=0.6,angle=90]{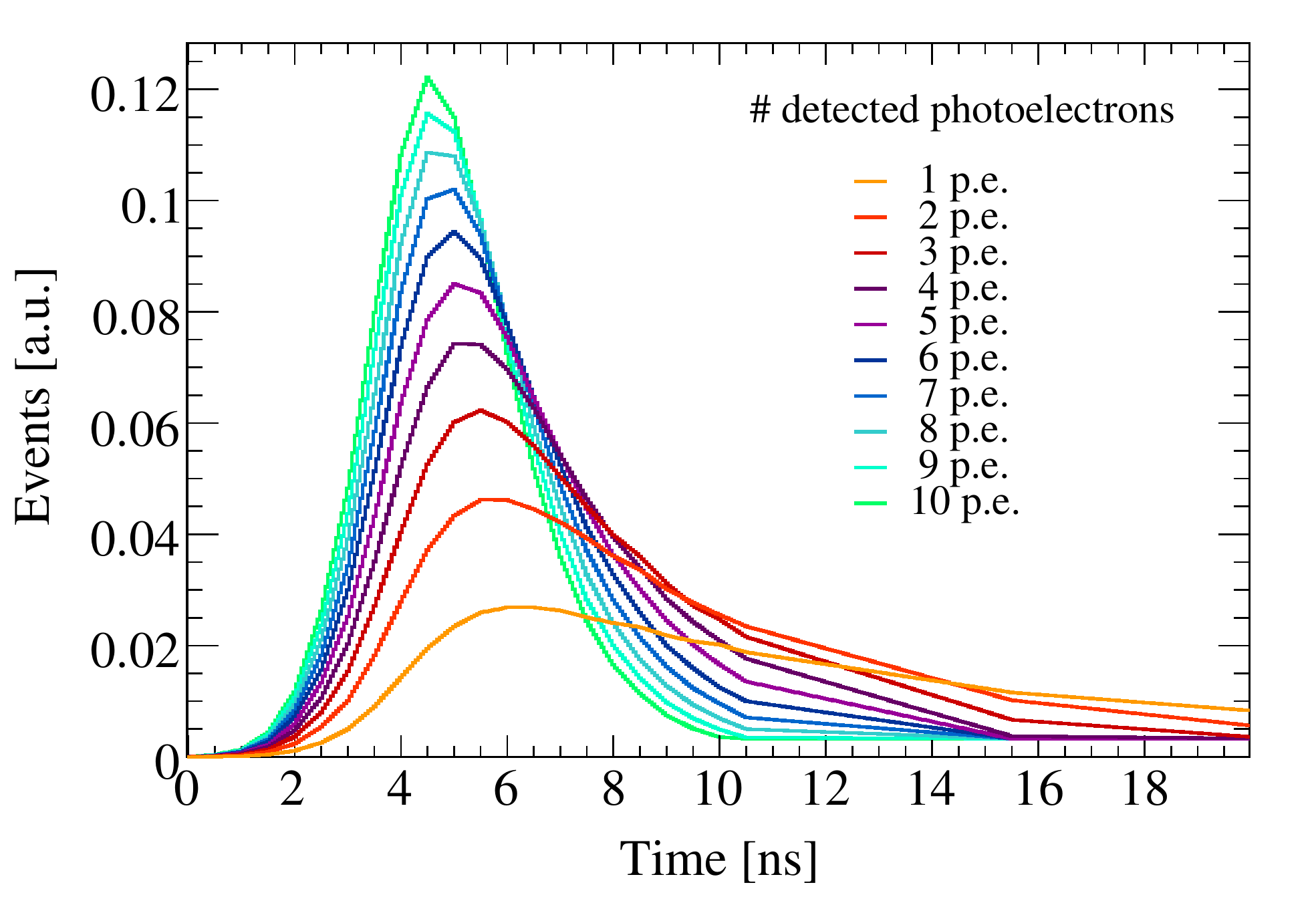}
\caption{\label{fig:pdf} 
Distributions of times at which the first photo-electron (p.e.) from an
event is detected by one of the Borexino PMTs, after time-of-flight subtraction.  By comparing the probability density functions ({\it pdf}s of events, where either a single
p.e. or multiple p.e. are detected, the time-of-arrival of the first p.e. is pushed to earlier time in the latter cases. These distributions 
are used as {\it pdf}s in the position reconstruction software algorithms.
}
\end{center}
\end{figure}

 The reconstruction of a physics event position $\vec r_0$ in Borexino 
 is based on the time distribution of the collected photons: 
 the algorithm considers for each photon its arrival time $t_i$ and 
 the position $\vec r_i$ of the PMT which 
 detected it, subtracts its time-of-flight $T^i_{flight}$, and compares
 the photon time distribution with the reference probability density function 
 ({\it pdf}) of the Borexino scintillator (Figure ~\ref{fig:pdf}). 
 The event position is calculated by maximizing the likelihood 
 $L_E(\vec r_0, t_0  \mid (\vec r_i, t_i))$ that the event occurs at 
 the time $t_0$ in the position $\vec r_0$ 
 given the measured hit time pattern  $(\vec r_i,t_i)$.

Note that the shape of the {\it pdf} depends on the charge $q$ collected 
at each PMT (Figure~\ref{fig:pdf}).
For events with energy deposition below 1\,MeV most of the PMTs work in the
single-photon regime, while at higher energies or close to the borders of 
the IV this is no longer the case, and the multi-p.e. effect has to be taken into account
to avoid bias in the reconstruction.

During the iteration of the minimization process the time-of-flight 
$T^i_{flight}$ of each single photon is calculated by
\begin{equation}
T^i_{flight}(\vec r_0, \vec r_i)  ={\frac { \mid \vec r_0 -\vec r_i \mid }{v_g}}
\label{eq:eq2}
\end{equation}
where  $\vec r_i$ are the coordinates of the PMT that has detected 
the $i$-$th$ photon. The parameter
$v_g$ corresponds to the group velocity of the wave packet emitted 
in the scintillation event  
\begin{equation}
 v_g  =
 {\frac {c}{ n- \lambda \cdot dn/d\lambda}}
\label{eq3}
\end{equation}
In the region of interest for Borexino, between approximately 350 and 600\,nm,
the variation of $n$ as function of $\lambda$ is relatively small 
($\simeq$3\%), however it has a strong impact on the effective velocity of 
the propagating scintillation wave packet, as can be seen from Equation \ref{eq3}. 
This has to be taken into account in the reconstruction algorithm by using 
an effective index of refraction
$n_{eff}$ significantly larger ($n_{eff}$=1.68) than the index of refraction 
 of pseudocumene measured at 600\,nm ($n_{\rm{PC}}$=1.50).
Calibration data were crucial in determining the exact value of $n_{eff}$ 
needed for a correct reconstruction of the event positions.
Note that prior to the calibrations the position reconstruction code was 
tuned using contaminants uniformly distributed in the scintillator such 
$^{14}$C and $^{222}$Rn. At that time, the systematic uncertainty on the 
position reconstruction was the dominant contribution to the total uncertainty budget
for the $^7$Be neutrino measurement \cite{borex_be7_2}.

\subsubsection{Position Reconstruction Using Radioactive Sources}
\label{sec:PRC}

The $^{222}$Rn and $^{241}$Am$^9$Be source data were used to estimate the quality and possible systematics of the position reconstruction
performance. As pointed out in Section \ref{sec:sources} the two sources 
were placed in $\sim$\,200 and $\sim$\,30 positions within the scintillator,
respectively. Figure~\ref{fig:srcspc} shows the energy spectra of the 
two sources in the variable {\it p.e.} described in Section~\ref{sec:ER}. 
Both sources cover most of the energy range of interest for the main Borexino neutrino analyses: the $^{222}$Rn source spectrum covers the region
of 0-3.2\,MeV suitable for the $^7$Be, {\it pep}, and CNO neutrino analyses. The prompt and delayed spectrum from the $^{241}$Am$^9$Be source measurement
is ideal for e.g. $^8$B solar neutrino and for geo-neutrino studies.

\begin{figure}[htbp]
\begin{center}
\includegraphics[scale=0.6]{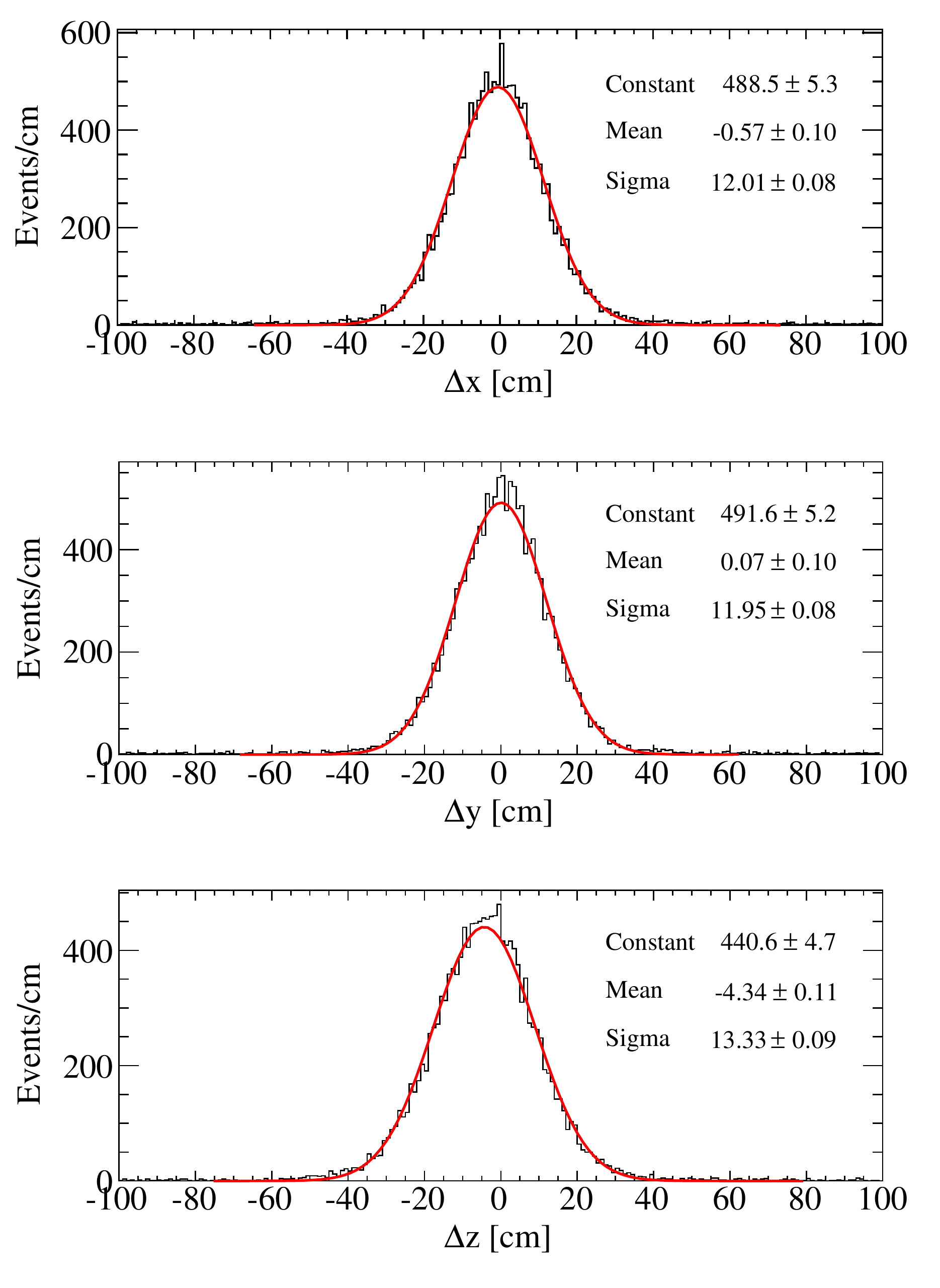}
\caption{\label{fig:Rn_example} Difference between the coordinates x, y and z reconstructed by software algorithms versus the nominal ones using
the CCD camera calibration system. The events used for the comparison are $^{214}$Po events emitted from a $^{222}$Rn calibration source placed
at the center of the detector.}
\end{center}
\end{figure}

As already mentioned in Section \ref{sec:cal_loc_sys}, the nominal source 
position was independently measured by the CCD camera system with an uncertainty of $\pm$0.6\,cm at 1$\sigma$.
Figure \ref{fig:Rn_example} shows the difference between the
reconstructed and nominal position for $^{214}$Po events from a 
$^{222}$Rn source measurement in the detector center:
the resolution of the distribution for these events ($\sim$700\,keV
electron equivalent) is 12\,cm ($\sigma$) for $x$ and $y$ and 13\,cm for $z$.

\begin{figure}[htbp]
\begin{center}
\includegraphics[scale=0.6,angle=90]{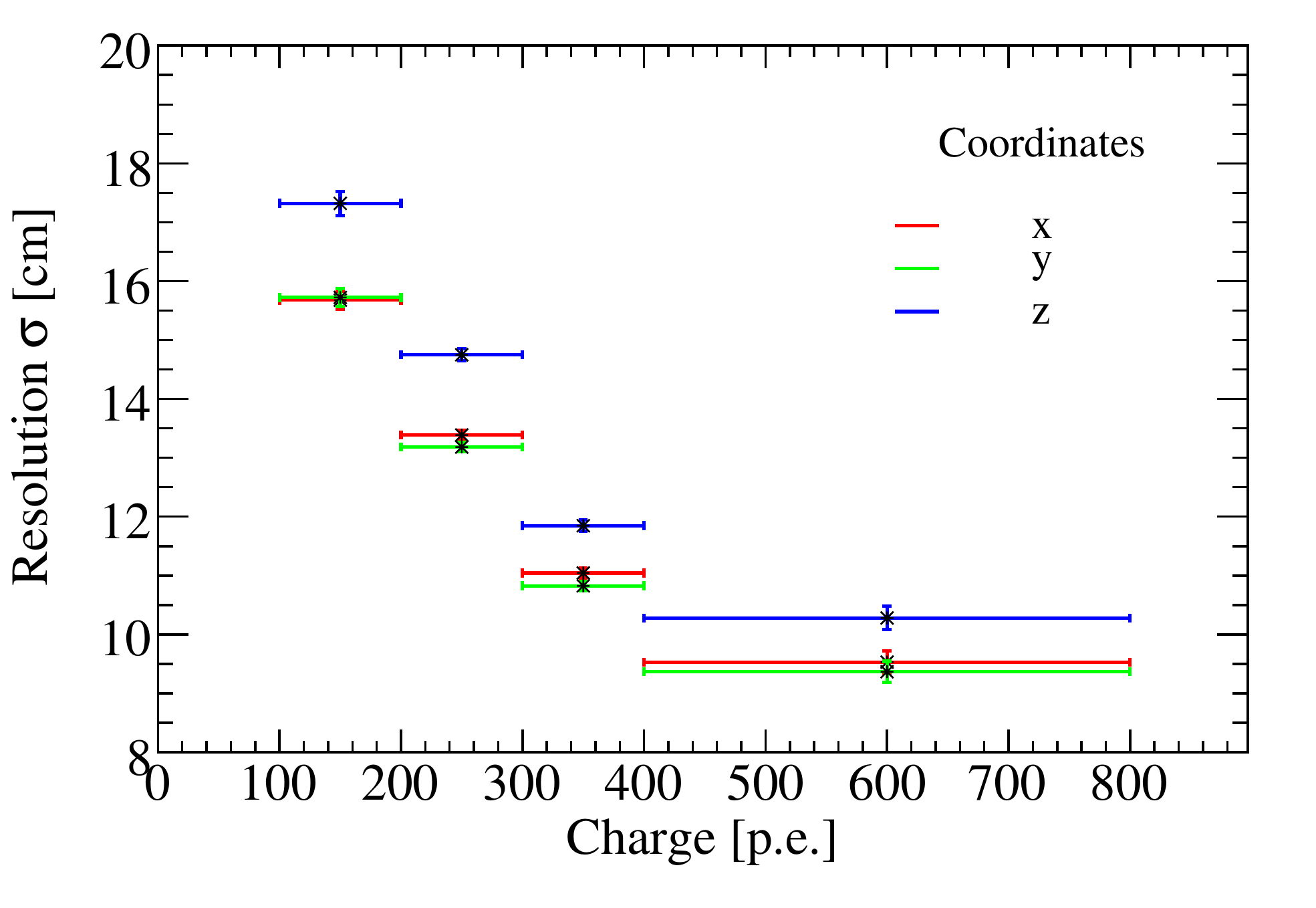}
\caption{\label{fig:Radon_resol} Resolution $\sigma$ for the three coordinates $x$, $y$ and $z$ as function of the collected charge (unit: number of photoelectrons). The results
are based on calibration data from the $^{222}$Rn and the $^{241}$Am$^9$Be source placed in the detector center.}
\end{center}
\end{figure}

The dependence of the position reconstruction resolution from the deposited
energy is shown in Figure \ref{fig:Radon_resol} for events reconstructed at the center:
the resolution for the $x$ and $y$ coordinates ranges from 15\,cm at $\sim$150\,p.e. to 9\,cm at $\sim$500\,p.e.
The resolution of the $z$ coordinate is $\sim$2\,cm worse, since
the PMT coverage in the $z$ direction is reduced.

\begin{figure}[htbp]
\begin{center}
\includegraphics[scale=0.6]{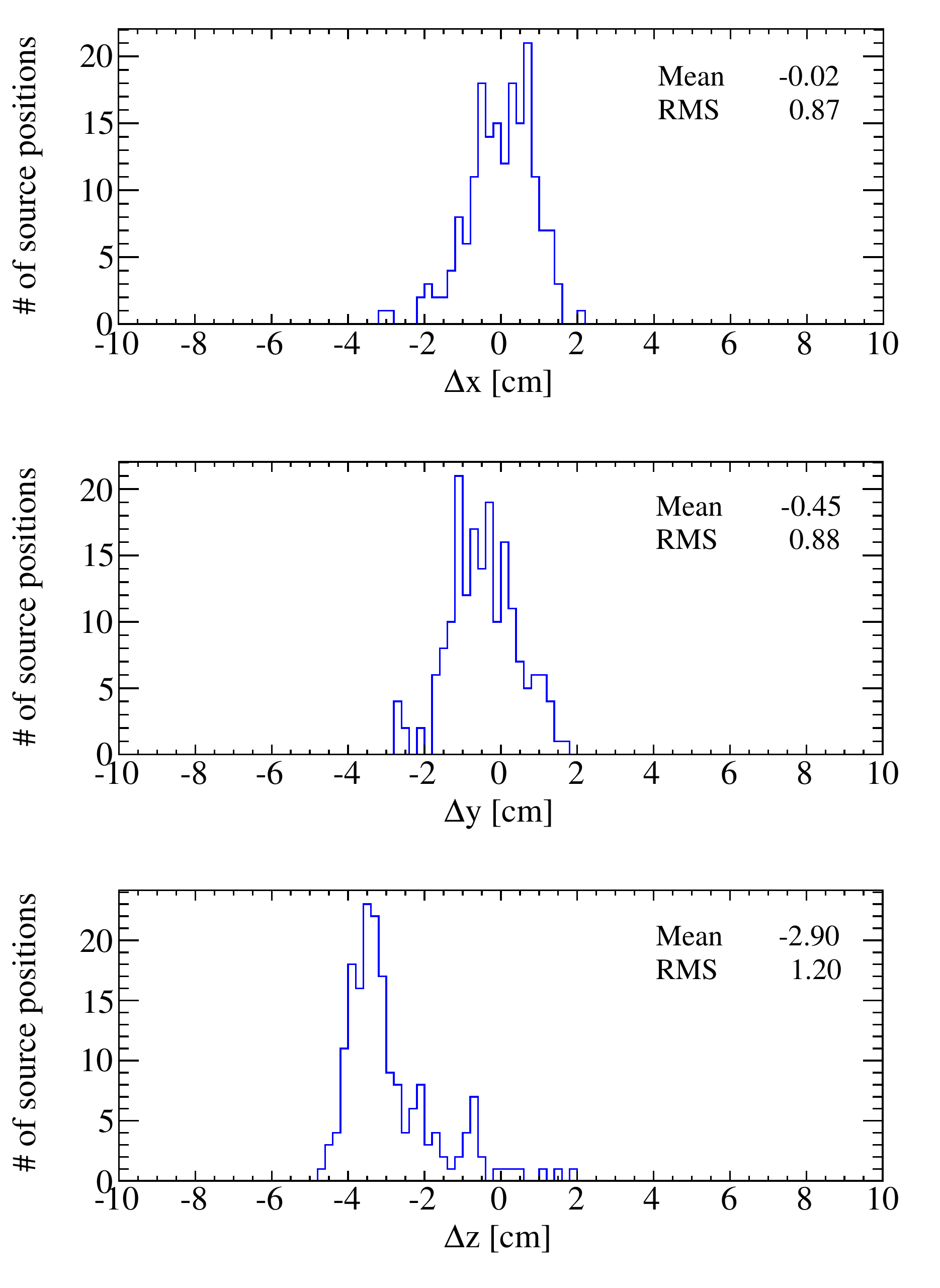}
\caption{\label{fig:xyz_rn} The $^{222}$Rn source was deployed in 182 different positions within the Borexino scintillator. For each
position the mean value of the event positions reconstructed by software algorithms as well as by the CCD camera calibration system was
determined. The plot depicts the difference between these two values
for the coordinates $x$, $y$ and $z$.}
\end{center}
\end{figure}

In order to investigate possible systematics associated with the position 
reconstruction, the reconstructed 
and nominal positions were compared for all  $^{222}$Rn source positions.
Figure \ref{fig:xyz_rn} depicts the difference between the mean value of the reconstructed coordinate ($x,y,z$)
and the corresponding nominal value for all available $^{222}$Rn source data.
The coordinates $x$ and $y$ are well reconstructed, i.e., 
the distribution is centered approximately on 0 with a sigma of 
 $\sim$1.5\,cm and tails up to 3\,cm maximum. The small bias (-0.45\,cm)
 in $y$ direction is negligible.
Note that the CCD reconstruction uncertainty of 0.6\,cm is included in the widths of the
distributions in Figure \ref{fig:xyz_rn}, and is not disentangled here.
A more significant bias of approximately -3\,cm in the reconstructed $z$ coordinate is observed.
The origins of this effect have not yet been understood: it may be the result of a
small offset in the position of the PMT coordinate system.
Nevertheless, as pointed out in the following Section \ref{fv}
the observed $z$-shift negligibly contributes to the systematic uncertainty in Borexino physics results.  

\begin{figure}[htbp]
\begin{center}
\includegraphics[scale=0.6]{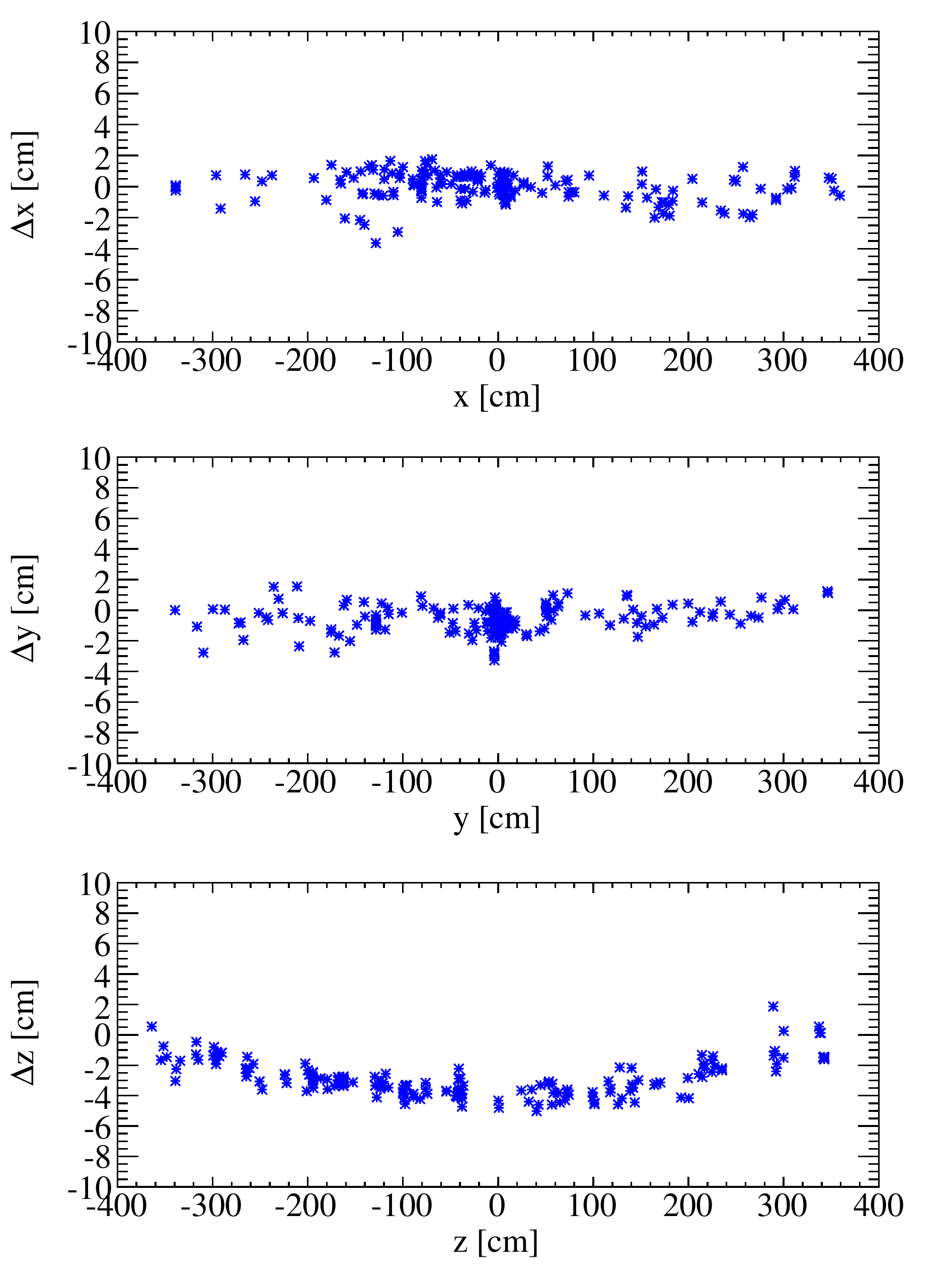}
\end{center}
\caption{\label{fig:possft} Difference $\Delta x$, $\Delta y$, $\Delta z$ between the mean values of the $^{214}$Po event position distributions reconstructed by software
algorithms and by the CCD camera calibration system as function of the coordinates $x$, $y$ and $z$.}
\end{figure}

The performance of the position reconstruction as a function
of position and energy was studied using both $^{222}$Rn and $^{241}$Am$^{9}$Be source
data. Figure~\ref{fig:possft} shows the difference between the
reconstructed and nominal coordinates as functions of the coordinate itself  
for $^{214}$Po events.
Biases in the reconstruction would show up in non perfectly flat 
distributions. The plot shows that the coordinates $x$ and $y$ are well reconstructed at all energies, while the $z$
coordinate is not flat near the detector poles and exhibits the already mentioned shift downwards. Irregularities in
the polar regions are due to shadowing effects of the IV plates that reduce the light collection.
Since typical fiducial volumes in Borexino exclude events reconstructed in the polar regions where the external radioactivity
contribution is higher, this effect has no impact on physics results.

\clearpage

\subsubsection{Fiducial Volume Determination and Systematics}
\label{fv}

The event position reconstruction is used to define a subregion of the
active volume called the fiducial volume (FV) via offline software cuts. The FV determination allows
us to efficiently exclude the
high radius background events emitted from detector materials surrounding the radiopure scintillator. 
Since these background events have different radial penetration depending on their energy, 
the FV definition depends on the 
energy region of interest for a given neutrino analysis.
The FV definition introduces an uncertainty in the target mass and, thus, in the neutrino rate determination. 
Calibration data were used to evaluate the systematic uncertainty associated 
with the FV selection for the different neutrino analyses.
For instance, the FV used in the $^7$Be neutrino analysis is defined 
by a radial cut $R$$<$3.021\,m 
and a $z$-cut of $|z|$$<$1.67\,m \cite{borex_be7_3}.
In order to estimate the systematic uncertainty on the fiducial mass, source data corresponding to 
positions at the border of the 
$^7$Be FV were selected. For this data set, the distributions of
$\Delta R$ and $\Delta z$, i.e. the difference between reconstructed and
nominal value of the radius and of the $z$ coordinate, were calculated. The FV systematic uncertainty 
was estimated by comparing the nominal value (86\,m$^3$) with the values obtained by varying 
$R$ and $z$ between the minimum and maximum $\Delta R$ and $\Delta z$.
Based on this, the FV contribution to the total systematic uncertainty budget of the $^7$Be neutrino rate
is +0.5\% and -1.3\%.
The systematic shift of 4\,cm in the $z$ direction (Section \ref{sec:PRC}) 
has a negligible impact on the selected FV, i.e. less than 0.01\%.

\subsection{Trigger efficiency}
\label{sec:trigger}

In addition to energy calibration and FV studies,
the  $^{85}$Sr source discussed in Section \ref{sec:sources}
 was also used for efficiency studies.
We know from an independent work performed with laser light, that the
trigger efficiency is $>$99.999\% for energies greater than $\sim$120\,keV.
This result was obtained exploiting the PMT equalization system 
\cite{PMT_timing} which provides pulsed laser light of known intensity
simultaneously to all the photomultiplier tubes.
A similar study can be performed with a radioactive source of known activity.
An advantage of this method is that it allows us to study the detector efficiency
as a function of position, by deploying the source in critical points
within the scintillator, for example at the borders of the FV. 
$^{85}$Sr decays with a lifetime of $\tau$=93.54\,d under emission 
of a characteristic 514\,keV $\gamma$ ray. This energy is relevant in particular
for the solar $^7$Be neutrino analysis. 
The activity of the source was measured 
with a germanium detector and found to be (3.28$\pm$0.07)\,Bq (reference
date: June 18, 2009). 
The source was then deployed in eight positions, some of which at the periphery of
the FV used in the $^7$Be neutrino analysis and its activity 
was measured with the Borexino detector by performing a spectral fit.
A spherical cut of $R$=1\,m centered on the nominal source position was needed
to reduce background from contaminants distributed in the scintillator volume,
such as $^{210}$Po.  
The detector efficiency was estimated by comparing the measured activity 
with the nominal one and was found to be consistent with unity in all positions,
showing no loss at the border of the Fiducial Volume.
From this analysis we find
$\epsilon$(E=514\,keV)=1.05$\pm$0.03\,(stat)\,$\pm$0.03\,(syst)
where the systematic error includes the uncertainty 
on the nominal activity of the source and the
systematic error associated with the cut and fit procedure applied to extract
the $^{85}$Sr source signal from data.

{\subsection{Vessel Shape Analysis}}
\label{sec:IVshape}

The inner vessel is one of the most fragile component in the 
design of the Borexino detector. This very thin 
nylon film separates two massive, and close-to-equal in density 
liquids and needs to be monitored carefully. 
The calibration systems discussed in Sections~\ref{sec:int} and~\ref{sec:ext_calibr_system} provide two independent methods to determine the
vessel shape\footnote{A third method for determining the vessel shape exploits 
the $^{210}$Bi contamination of the vessel and will not be discussed here.}.
These tools became particularly important in identifying a significant 
deformation of the vessel in May 2009 which 
was the result of a small leak of the active scintillator into the buffer
\footnote{After the discovery of the leak, the vessel has been refilled 
with scintillator and the leak rate has been reduced to tolerable levels
by diminishing the difference in density between the scintillator and the 
buffer liquids.}. 

\begin{figure}[htb] 
\centering
\includegraphics[scale=0.6, angle=90]{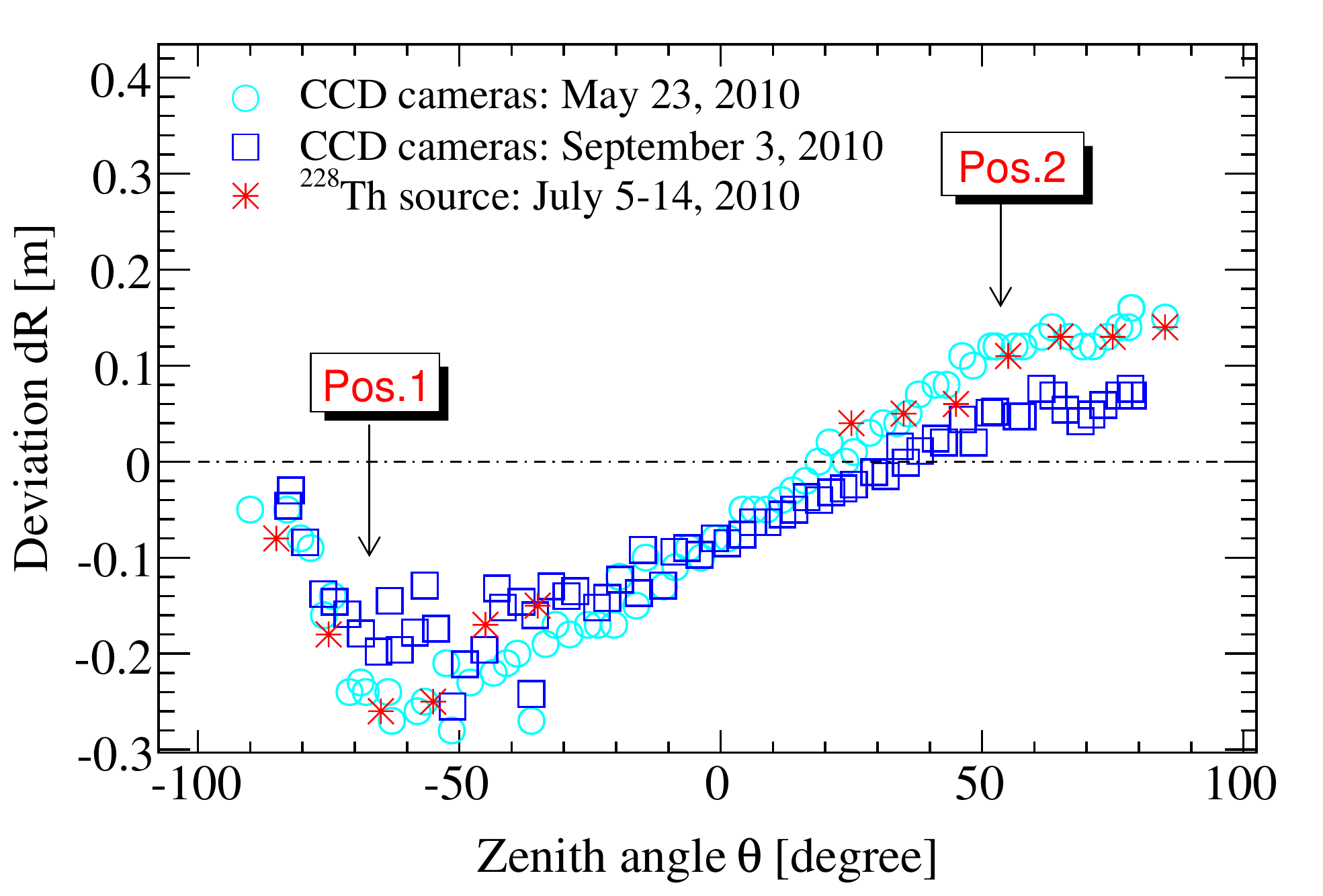}
\caption{Comparison of the vessel shape reconstruction using the CCD camera photos and the external $^{228}$Th source.
The CCD camera images were taken on May 23 and September 9, 2010,  while the {\it external calibration} data covers the period
between July 5-14, 2010. The $^{228}$Th source was placed only on the two zenith angles 53.5$^\circ$
and -67.2$^\circ$, thus not all the vessel was irradiated. Both methods lead to similar results,
where the nominal vessel radius $R$=4.25\,m is deformed by as much as $dR$=$\pm$
0.3\,m close to the poles.}
\label{fig:vessel_shape}
\end{figure}

The first of the two methods used for the inner vessel shape reconstruction 
is given by the CCD camera system. 
For this purpose the high voltage of all PMTs has to be turned off and the 
halogen-lights switched on. Under these conditions the vessels are easily visible to the CCD cameras in 
most locations due to small amount of light reflected from the vessel surface
facing the camera housings. Photos are taken from all the cameras to visually mark ($r,\vartheta$) points
on the edge of the vessel. Each of the points can be used to project a ray back to the camera which took the picture. 
This ray is, by definition, also a tangent to a sphere centered on the detector's origin. To the first order, the radius 
of this tangent point is the radius of the vessel at the given $\vartheta$. The vessel profile obtained with this method can be
seen in Figure~\ref{fig:vessel_shape} (empty circles and empty squares) which shows the deviation $dR$ from the nominal 
value of the vessel radius ($R$=4.25\,m) as a function of the zenith angle $\vartheta$.

The second technique relies on $\gamma$ ray tomography of the nylon vessel using the external $^{228}$Th source
(see Section~\ref{sec:ext_calibr_system}): the nylon vessel represents the transition region between scintillator and
the largely non-scintillating buffer liquid. By irradiating the vessel from the outer detector region, it is therefore possible to
locate the highest-radius visible events in each direction from the detector center. For the extraction of the angular-dependent radial
information the following method is applied. First, the zenith angle $\vartheta$ and the relative distance from detector center is calculated for all events induced by
the external $\gamma$ source. Then the radial distribution of events lying in separated slices of $\vartheta$ are plotted.
An example for events collected during the second external calibration campaign with d$\vartheta$=(-50$^{\circ}$;-55$^{\circ}$) is shown in Figure \ref{fig:extBack_radialDist}.
Finally, the radial distributions obtained for the different slices $d\vartheta$ are fitted with a convolution of the detector 
response and four functions described in more detail in Section~\ref{sec:ext_calibr_radial-distr}.
It turns out that only a short calibration time of one day is required in order to extract the vessel shape information with
the 5\,MBq$^{228}$Th source: the irradiation of the inner vessel from four almost equidistant source positions around the SSS, i.e. two in the equatorial region and two in the
upper and lower polar region, is sufficient for a precise extraction of the vessel shape information. 
Herein, the fit uncertainty of the angular-dependent radius is typically around 2\,cm. Figure~\ref{fig:vessel_shape} shows the vessel profile obtained with this
method (red stars) superimposed to the ones obtained with the CCD camera system before (blue circles) and after (blue rectangles) the first external calibration campaign. The results of the two methods are in good agreement.

Once the $(r,\vartheta)$ shape profiles have been recorded, an averaged curve can be computed to perform a numerical
integration to determine the volume of the inner vessel. A $\pm 2$\,cm error on the position of an item in the
detector translates to an uncertainty of approximately $1.5\%$ on the \emph{absolute} volume for a typical set of
photos. Relative errors between two sets of photos are lower -- at the level of $\pm$1\,m$^3$ over the volume
of the inner vessel, a $0.3\%$ accuracy.

\subsection{External Background}
\label{sec:ext-bkgr}

\subsubsection{Energy Spectrum of the External Background}
\label{sec:ext_calibr_spectra}

\begin{figure}[htb]
\centering
\includegraphics[scale=0.6, angle=90]{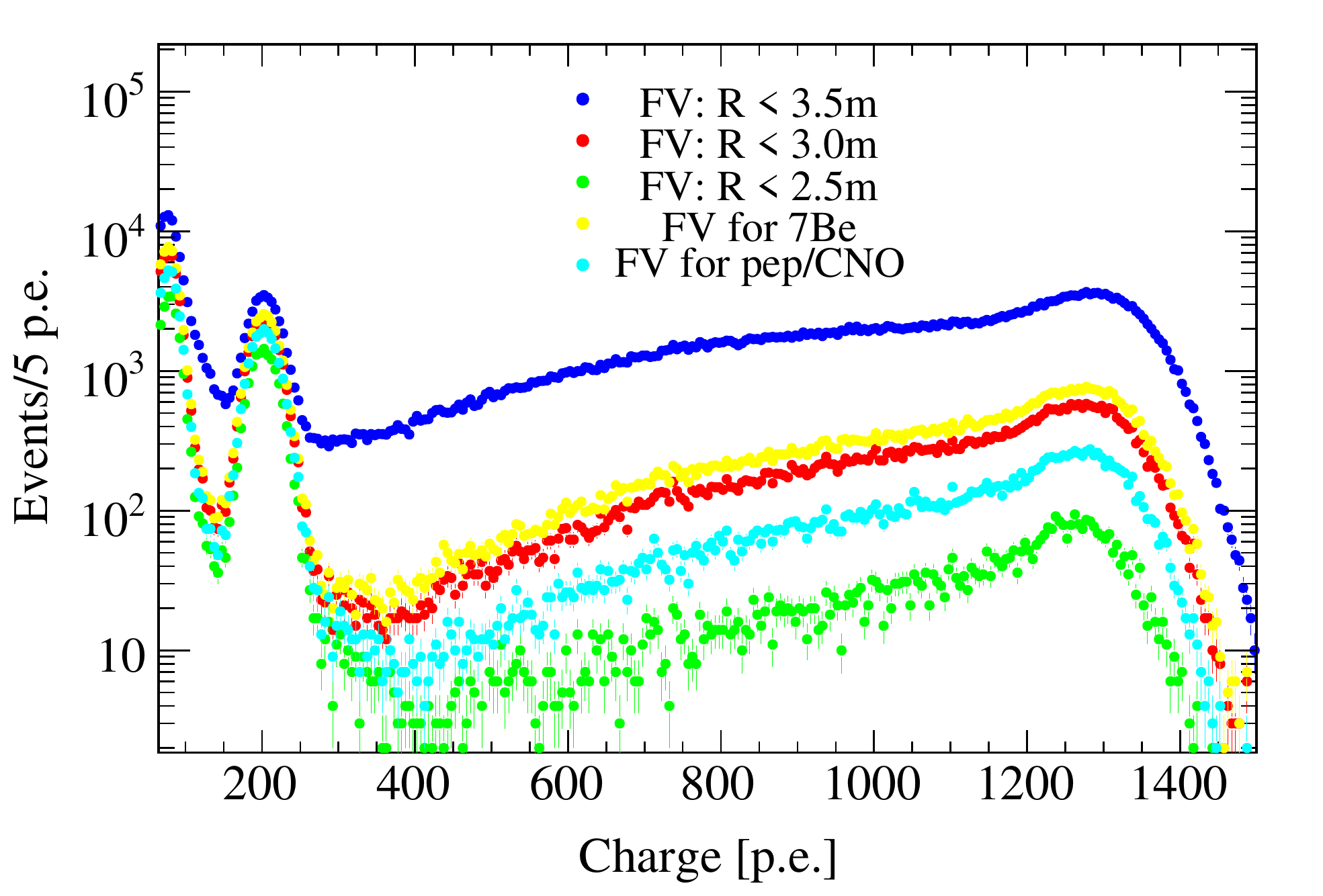}
\caption{Energy spectra of external $\gamma$ rays emitted by the $^{228}$Th source for different fiducial volumes. All data acquired during the
second {\it external calibration} campaign in different source positions were superimposed. The spectra from contaminants such as $^{14}$C and
$^{210}$Po decays were not subtracted and are still visible at lower energies.}
\label{fig:external-gammas_spectra}
\end{figure}

One goal of the external source calibrations was the 
determination of the energy spectrum of the external
$\gamma$ rays originating from the outer parts of the Borexino detector. This background is dominated by 2.615\,MeV $\gamma$ rays from
$^{208}$Tl decays in the detector materials surrounding the organic liquids. The
corresponding energy spectrum is expected to have a radial dependence. Since Compton-scattered 2.615\,MeV $\gamma$ rays have a reduced probability to reach the inner core
of the scintillator, the energy spectra of external background events reconstructed within small fiducial volumes (FV)
around the detector center have a suppressed Compton continuum and vice versa. However, the exact shape of the Compton
continuum for a given FV is {\it{a priori}} not known. The shape depends on the detector dimensions and structure
as well as on the materials used. Thus, the shape cannot be deduced analytically with a satisfactory precision.
The calibration with the external $^{228}$Th source allowed us to reproduce the energy spectrum of external 2.615\,MeV
$\gamma$ rays in Borexino. The obtained spectra for five different concentric FVs are shown in Figure
\ref{fig:external-gammas_spectra}. Three FVs are spherical and include all events reconstructed within a
radius $R$, whereas the FVs used for the solar $^{7}$Be as well as {\it pep} neutrino analyses have a
radial cut of $R$$<$3.021\,m and $<$2.8\,m, and a vertical $z$-cut of $|z|$$<$1.67\,m and -1.8\,m $<$z$<$2.2\,m, respectively. Rates and spectral shapes
are different. In the cases of spherical FVs with $R$$<$2.5\,m and $<$3.0\,m the rates vary from 1.3 to
4.4\,events/(day$\times$100\,ton) in the energy region [0.2,3.0]\,MeV. The fraction of events lying
below 2.0\,MeV are $\sim$20\% and $\sim$30\%, respectively. Below 0.2\,MeV the external background contribution
is negligible in both FV cases.

On the one hand, the deduced experimental energy shape is used to test the Borexino Monte Carlo code \cite{MC_paper}.
On the other hand, the data are used to study the position dependence of the energy reconstruction. Finally, the obtained external
background profile can be used in global spectral fits as an experimentally known component.

\subsubsection{Radial Distribution of the External Background}
\label{sec:ext_calibr_radial-distr}

\begin{figure}[htb]
\centering
\includegraphics[scale=0.6, angle=90]{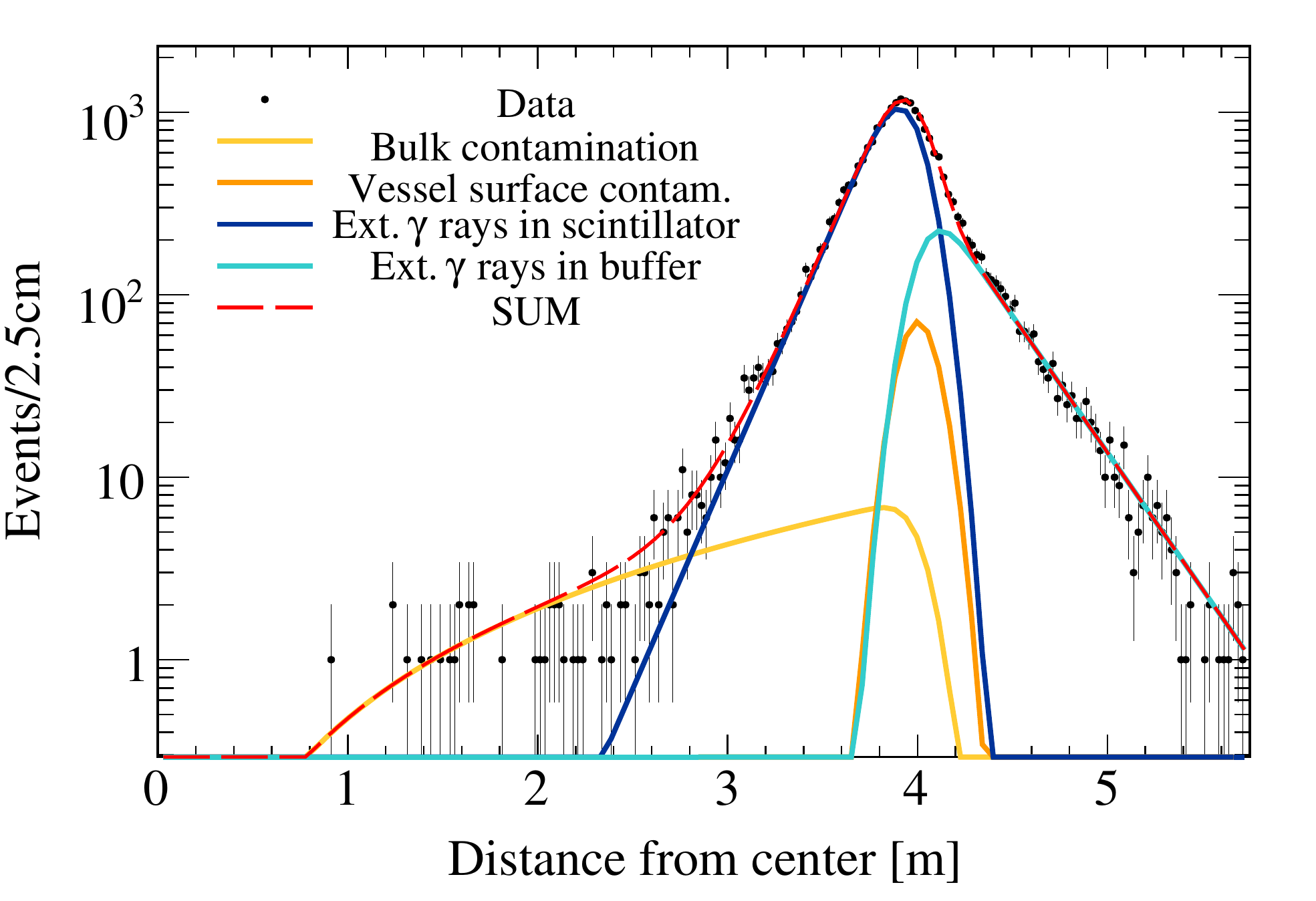}
\caption{Radial distribution of $\gamma$ rays induced by the $^{228}$Th 
source placed at 6.85\,m distance from the center of detector. The fit includes four components convoluted with the detector response. More details are given in the text.}
\label{fig:extBack_radialDist}
\end{figure}

The definition of the optimal fiducial volumes in Borexino requires a precise knowledge
of the spatial distribution of external background events.
To a first approximation, given the spherical symmetry of the detector and 
since most of the external
background sources are located near the SSS surface, the 
distribution of background events depends on radius only.
Thus, the radial dependency of the external background in Borexino has been 
studied by means of the externally placed $^{228}$Th source.
The radial distribution of such events is shown in 
Figure \ref{fig:extBack_radialDist}. 
The distribution of events reconstructed in the scintillator and buffer was fitted taking into account four components.
The first component corresponds to a volumetric function. It describes the bulk contamination given mainly by $^{14}$C and
$^{210}$Po decays that are isotropically distributed in the scintillator and represent an intrinsic background in the external calibration data.
The second fit component represents the $\beta$ contamination on the nylon surface of Inner Vessel.
The third distribution is an exponential function. It reflects the $\gamma$ rays that are emitted by the external calibration source and that are able to penetrate the scintillator volume
towards the detector center.
The fourth component is generated by the external source $\gamma$ rays that deposit most of their energy in
the buffer region where the emission of scintillation light is suppressed by more than one order of magnitude. The larger the distance of the reconstructed light baricenter from the Inner Vessel the
lower its probability to be detected. As it turns out, this attenuation behavior can be well described by an exponential function.
Finally, the weighted sum of all four components is convoluted with a Gaussian detector response function.
The fit procedure allows to extract several parameters such as the radius of the 
Inner Vessel. An application of it has already been shown in Section \ref{sec:IVshape}. 
Moreover, the position resolution and 
the attenuation length for 2.615\,MeV $\gamma$ rays in PC are determined 
to be $\sim$11\,cm and $\sim$25\,cm, respectively. 
The position resolution is in agreement with the one obtained from the {\it
internal calibration} data (see Section~\ref{sec:results_position-reco-fv}).
The attenuation length is in agreement with the reference values provided by the National Institute of Standards and Technology (NIST) 
database \cite{nist}. 

\subsubsection{Total Thorium Activity in the Outer Detector Components}
\label{Th228-activity-from-detector}

Materials used for the construction of Borexino were screened prior to
installation by means
of highly sensitive detection techniques in order to verify their
level of radiopurity. For practical reasons, only subsamples of larger 
batches were screened.
For example, this was the case for the stainless steel of the SSS, for the
PMTs and the LCs \cite{borex_material_screening}.
The estimation of the total radioactivity in the final detector assembly 
is based on the assumption that the subsamples are representative for the whole batches.

\begin{table*}[htb]
\begin{center}
\begin{tabular}{ccccccc}
\hline
\hline
Detector		&	$^{232}$Th		&	$^{238}$U		&	$^{40}$K$_{nat}$	&	Distance from&	Mass		\\
component		&	[g/g]			&	[g/g]			&	[g/g]			&	center [cm]	&	[kg]	\\
\hline
SSS			&	3$\times$10$^{-9}$	&	7$\times$10$^{-10}$	&	7$\times$10$^{-8}$	&	6.85		&	7$\times$10$^{4}$	\\
PMT			&	3.3$\times$10$^{-8}$	&	6.6$\times$10$^{-8}$&	2$\times$10$^{-5}$	&	6.55		&	2$\times$10$^{3}$	\\
Light concentrators	&	1.8$\times$10$^{-7}$	&	1$\times$10$^{-9}$	&	1$\times$10$^{-5}$	&	6.45		&   1.2$\times$10$^{3}$	\\
PC buffer		&	1$\times$10$^{-15}$	&	1$\times$10$^{-15}$	&	1$\times$10$^{-12}$	&	4.25-6.85	&	1.04$\times$10$^{6}$\\
O.V. Steel endcaps	&	2$\times$10$^{-9}$	&	1$\times$10$^{-9}$	&	7$\times$10$^{-8}$	&	5.75		&	1.5$\times$10$^{1}$	\\
I.V.-O.V. Nylon pipe&	5$\times$10$^{-11}$		&	5$\times$10$^{-11}$	&	7$\times$10$^{-7}$	&	--			&	4.2					\\
I.V. Nylon endcaps	&	5$\times$10$^{-11}$	&	5$\times$10$^{-11}$	&	7$\times$10$^{-7}$	&	4.25		&	1.2$\times$10$^{1}$	\\
Hold down ropes		&	5$\times$10$^{-11}$	&	5$\times$10$^{-11}$	&	1$\times$10$^{-6}$	&	4.25		&	4.5					\\
Nylon bag		&	4$\times$10$^{-12}$	&	2$\times$10$^{-12}$	&	1$\times$10$^{-8}$	&	--			&	3.2$\times$10$^{1}$	\\
\hline
\hline
\end{tabular}
\caption{\rm{Detector components contributing to the external background outside the {\it{Inner Vessel}}. The mass fractions of $^{232}$Th,
$^{238}$U and $^{40}$K, the total mass of the component and its average distance from the center of detector is included \cite{borex_material_screening, LC_Dissertation, light_concentrators}.}}
\label{tab:ext-background-det-components} 
\end{center}
\end{table*}

The measurement performed with the $^{228}$Th source allows us to measure directly
the total activity of thorium $A_t$ in the outer detector components for the
first time. This independent measurement can be compared with the overall estimation 
obtained from the material screening results. 
The method is based on a comparison of count rates in normal data and in 
{\it external calibration} data.
For a spherical fiducial volume (FV) with a radial cut at $R$$<$3.0\,m the 
external background rate in normal data is estimated by a spectral fit, 
using the shape derived experimentally from {\it external calibration} data.
The rate is found to be (4.4$\pm$1.0)\,events/(day$\times$100\,ton) in the energy region [0.2,3.0]\,MeV.
For the same FV conditions and the same energy range, the rate of events induced by the 
$^{228}$Th source is measured to be $\sim$3600\,events/(day$\times$100\,ton).
Since the radial position of the source and its activity is well known with an
uncertainty of $\pm$3\,cm and 6\%, respectively, it is
possible to deduce the overall activity of the external detector components for the assumption that all of them are
located at a distance $R$=6.85\,m from the detector center. This is possible by rescaling the rate observed in normal data
with respect to the one measured on calibration data. This leads to
 \begin{equation}
A_t(^{228}\mathrm{Th}) = (6.1\pm1.4) \,\,\mathrm{kBq}
\end{equation}
This result is compared with the expectation from the material screening. 
The main contribution to the external background measured in
Borexino comes from the SSS, the PMTs and the LCs.
 Their distances from the detector center are 6.85\,m, 6.25-6.65\,m and 
 6.55-6.85\,m, respectively. 
 Moreover, secular equilibrium of the $^{232}$Th decay chain is assumed. 
 Then, according to Table \ref{tab:ext-background-det-components} the total $^{228}$Th activities
 of the three aforementioned components were measured to be 0.3\,kBq (SSS), 0.1\,kBq (PMTs) and 
 0.3\,kBq (LCs), respectively \cite{borex_material_screening,LC_Dissertation,light_concentrators}. 
The corresponding equivalent activity assumed to be due to a single component
located at a distance 6.85\,m from the center of the detector is
 \begin{equation}
A_{t,MS}(^{228}\mathrm{Th}) \sim{4.7} \,\,\mathrm{kBq}
\end{equation}
This value is in good agreement with the result derived from the comparison of Borexino 
normal and {\it external calibration} data.

\section{Conclusions}

The Borexino calibration system has been presented.
It includes an internal and an external part, the former used to
deploy radioactive sources inside the scintillator, and the latter designed to
place $\gamma$ sources outside of it close to the stainless steel sphere (SSS) and the photomultiplier tubes (PMTs).

The calibration hardware, sources and procedures were reviewed. 
Several  internal and external calibration campaigns were successfully 
performed in the period between 2008 and 2011. In total,
twelve different $\alpha$, $\beta$, and $\gamma$ sources were deployed in 295 
positions in the highly radio-pure Borexino scintillator
without introducing any detectable contamination. 
Moreover, a 5\,MBq $^{228}$Th $\gamma$ source was placed in
ten positions close to the surface of the SSS in order to
study the external $\gamma$ ray background originating from the outer detector components.

Several important results were achieved.
The position reconstruction algorithm was calibrated at a 
precision of a few percent using the  internal calibration
data. As a result, the systematic uncertainty of the fiducial volume 
and thus of the solar neutrino rate could
be drastically reduced \cite{borex_be7_3}. 
The  internal calibration campaigns also provided important information 
regarding the energy response of the Borexino detector, opening the possibility
for a precise determination of the absolute energy scale. 
Moreover, they allowed us to study the energy reconstruction uniformity
within the standard fiducial volumes, as well as asymmetries in 
position and energy reconstruction due to, for instance, the non-uniform 
distribution of operational PMTs. 
The energy spectra and spatial distributions of reconstructed calibration 
events were also important 
to test and validate the Borexino Monte Carlo code \cite{MC_paper}.
The  external calibration campaigns allowed us to study both the spectral 
and the radial shape of the external background in Borexino.
Finally, the  external calibration data allowed us to determine detector properties
such as the shape of the nylon vessel containing the scintillator 
as well as the thorium content in the external detector
components.

The calibration data had a crucial role in the success of Borexino
and will have a strong impact on its future physics results. 
New calibration campaigns will be carried out at the end of the recently begun second
phase of the Borexino experiment.
Besides further improvement in the precision of the physics results,
the new calibration campaigns will allow a long-term stability study of the
detector response and of the scintillator properties over a period of more than five years.
This is of major importance for future rare-event physics experiments that will be based
on organic liquid scintillators as a detection medium.

\section*{Acknowledgements}
\label{sec:acknowledgements}

This work was funded by INFN and MIUR PRIN 2007 (Italy), NSF (USA),
BMBF, DFG, and MPG (Germany), NRC Kurchatov Institute (Russia), and MNiSW (Poland). We gratefully acknowledge the generous support of the Laboratori
Nazionali del Gran Sasso. We thank the Swiss Paul Scherrer Institut in Villigen, the German Institut f{\"u}r Kernchemie at Johannes-Gutenberg University in
Mainz and the German Physikalisch-Technische Bundesanstalt in Braunschweig for their scientific cooperation in the construction and characterisation of the
custom-made $^{228}$Th source.

\bibliography{calibration_paper}

\end{document}